\documentclass[bibyear]{aa}  
\usepackage{hyperref}
\usepackage{graphicx}
\usepackage{amsmath}
\usepackage{enumitem}
\usepackage{natbib}
\usepackage{booktabs}
\usepackage[normalem]{ulem}
\usepackage[table,xcdraw]{xcolor}
\usepackage{lineno}
\usepackage{subcaption}
\usepackage{lipsum}
\usepackage{acronym}
\acrodef{SDE}[SDE]{Stochastic Differential Equation}
\acrodef{GCR}[GCR]{Galactic Cosmic Ray}
\acrodef{CUDA}[CUDA]{Compute Unified Device Architecture}
\acrodef{GPU}[GPU]{Graphic Processing Unit}
\acrodef{CPI}[CPI]{Charged Particle Instrument}
\acrodef{CIR}[CIR]{Corotating Interaction Region}
\acrodef{HMF}[HMF]{Heliospheric Magnetic Field}
\acrodef{SEP}[SEP]{Solar Energetic Particles}
\acrodef{VLIS}[VLIS]{very local interstellar spectrum}
\acrodef{LIS}[LIS]{local interstellar spectrum}
\acrodef{TPE}[TPE]{Transport Equation}
\usepackage{txfonts}

\begin{document} 


\title{Numerical and experimental evidence for a new interpretation of residence times in space}

\titlerunning{Evidence for a new interpretation of residence times}


   \author{A. Vogt\inst{1}
          \and
          N. Eugene Engelbrecht\inst{2}
          \and
          B. Heber\inst{1}
          \and
          A. Kopp\inst{2,3}
          \and
          K. Herbst\inst{1}
          }

   \institute{Institut f\"ur Experimentelle und Angewandte Physik, Christian-Albrechts Universit\"at zu Kiel,
Leibnizstra\ss e 11, D-24118 Kiel, Germany\\
              \email{vogt@physik.uni-kiel.de}
         \and
             Centre for Space Research, North-West University, 2520 Potchefstroom, South Africa
         \and
            Theoretische Physik IV, Ruhr-Universit\"at Bochum, Universit\"atsstr. 150, 44801 Bochum, Germany
             }

   \date{}

 
  \abstract
  {} 
   {We 
   investigate the energy dependence of Jovian 
   electron
   residence times, which allows for a deeper understanding of adiabatic energy changes that occur during charged particle transport, as well as 
   of
   their significance for simulation approaches. Thereby we seek to further validate an improved approach to estimate residence times numerically by investigating the implications on 
   previous
   analytical approaches, and possible effects detectable by spacecraft data. }
   {Utilizing a propagation model based on a Stochastic Differential Equation (SDE) solver written in CUDA, residence times for Jovian electrons are calculated over the whole energy range dominated by the 
   Jovian electron
   source spectrum. We analyse the interdependences both with the magnetic connection between observer and the source as well as between the the distribution of the exit (simulation) times and the resulting residence times.}
   {We point out a linear relation between the residence time for different kinetic energies and the longitudinal shift of the 13 month periodicity typically observed for Jovian electrons and discuss the applicability of these findings to data.  Furthermore, we utilize our finding that the simulated residence times are approximately linearly related to the energy loss for Jovian and Galactic electrons, and develop an improved analytical estimation in agreement with the numerical residence time and the longitudinal shift observed by measurements.}
   {}


   \maketitle
%

\section{Introduction}

Since \cite{Pyle1977} confirmed the Jovian magnetosphere as a dominant point-like source of low-MeV electrons, these so-called Jovian electrons have been utilised as test particles in order to investigate charged particle transport in the inner Heliosphere. The decentral position of the Jovian source within the \ac{HMF} thereby offers the possibility to distinguish between parallel and perpendicular diffusion. Many computational studies of Jovian electron transport \cite[see, e.g.,][]{Chenette77,Conlon1978,Fichtner2000, Ferreira2001,Kissmann2003,Ming07,Sternal2011} have yielded valuable insights into the diffusion tensor have utilised Jovian electrons as test particles by taking advantage of Jupiter's position in the HMF, under the assumption that electron transport during periods of good magnetic connection between this source and the Earth would be dominated by the diffusion coefficient parallel to the HMF, and be dominated by the perpendicular diffusion coefficient during times of poor magnetic connection. With the advent of stochastic techniques for solving the \cite{parker1965} \ac{TPE} that describes the transport of these particles \cite[see, e.g.,][]{Zhang1999,Pei2010,Strauss2017}, it has become possible to study the residence times of these particles, and subsequently the influence of various transport processes on them \cite[e.g.][]{Florinski2009,strauss2011}. The present study aims to investigate, in particular, the influence of adiabatic energy changes on the residence times of Jovian electrons.

 In order to estimate the residence times of charged particles in the context of \ac{SDE} modelling, \cite{vogt2019} proposed a new approach with results more realistic than 
 the ones by
 previous attempts. This study will investigate further the energy dependence of residence times 
 regarding
 how they are related to adiabatic energy changes as modelled by the Parker \ac{TPE} in particular. Therefore, we will re-examine the suggested calculation of residence times with respect to the influence of adiabatic energy changes and investigate the relation between these two quantities. We will analyse the distribution of simulation or exit times as provided by the \ac{SDE} approach with respect to their contribution to the residence times.
 In the following, we also expand our considerations from the inter-dependence of residence times and adiabatic energy changes within the \ac{SDE} modelling scheme and discuss the consequences of our results for the understanding of diffusion approaches within the physics of charged particle transport in general. Finally we compare predictions based on our results with spacecraft observations obtained by SOHO-EPHIN.
 
 
 As \cite{vogt2019} focused on 6 MeV Jovian electrons, this work subsequently aims to expand the focus to energies covering the whole observed range of $E_{Jov}=[0.3,100.0]~$MeV as indicated by the Jovian source spectrum according to \cite{vogt2018}.
 The goal is to gain further insights as to the role that adiabatic energy changes play relative to diffusion to mathematically model charged particle propagation. 
 In order to do this, we implement a parallel and perpendicular mean free path independent of the particle's energy as discussed in Sec.~\ref{ssec:setup} by Eq.~(\ref{eqn:lambda_par}). 
 
 In that regard we will also discuss prior analytical attempts to estimate the residence times of \acp{GCR}, namely by \cite{parker1965} and \cite{OGallagher1975}. 
 In order to confirm our considerations
 , we discuss the possibility to 
 validate our results via aspects of
 spacecraft data caused by the corotation of the Jovian source relative to the observer during the particles propagation through interplanetary space. 
 By these means we aim to establish that the estimation for the residence time as proposed by \cite{vogt2019} is consistent both with the analytical treatment of the \ac{TPE} as well as with observations of Jovian electrons for the entire range of these particle's energies.

\section{Scientific Background}

\subsection{Jovian electrons}

Jupiter has been known as a source of an electron population dominating the inner heliosphere in the low-MeV range since the early 1970s. After being proposed by \cite{McDonald72}, based on a characteristic 13-month periodicity in electron counting rates obtained at Earth orbit, the \textit{Pioneer 10} flyby was able to prove the origin of this phenomenon 
as
the Jovian magnetosphere. \cite{Teegarden1974}, as well as \cite{Pyle1977}, showed that the 13 month periodicity was indeed linked to the synodic period of Jupiter, and that the source could be approximated to be point-like. 

These two unique qualities, the point-like nature of the source itself and its decentral position, led to ongoing efforts to model Jovian electron propagation, beginning with \cite{Conlon1978}. Since the decentral position causes changes in the magnetic connectivity between Jupiter and a possible observer, Jovian electrons are especially suited to serve as test particles to investigate parallel and perpendicular diffusion coefficients. Previous studies on this have been published by \citep[e. g.][and references therein]{Chenette77,Fichtner2000, Zhang-etal-2007}. Furthermore, a significant influence of drift effects on low-MeV electron transport has been ruled out both from a modelling \citep[e. g.][]{Potgieter1996,Burger2000,Ferreira2002} as well as from a theoretical perspective \citep[e. g.][]{Bieber1994, EB2010, EB2013}. 

The source spectrum itself remained a topic of research 
as well
, due to the general difficulties in measuring electron spectra, as detailed by \citep[e. g.][]{Heber2005, kuehl2013}. Based on both \textit{Pioneer} 10 flyby and Earth orbit data \citep[e. g.][amongst others]{Teegarden1974, Baker1976, Eraker1982, Ferreira2001} published suggestions on both the general shape of the spectrum and its strength. A recent estimation based on the \textit{Pioneer} 10 and \textit{Ulysses} flyby data, as published by \cite{Heber2005} in the meantime, has been given by \cite{vogt2018} and is utilized for this study.

\subsection{\ac{SDE} based modelling}
\label{ssec:SDE_modelling}

\begin{figure*}
    \centering
    \includegraphics[width=0.99\textwidth]{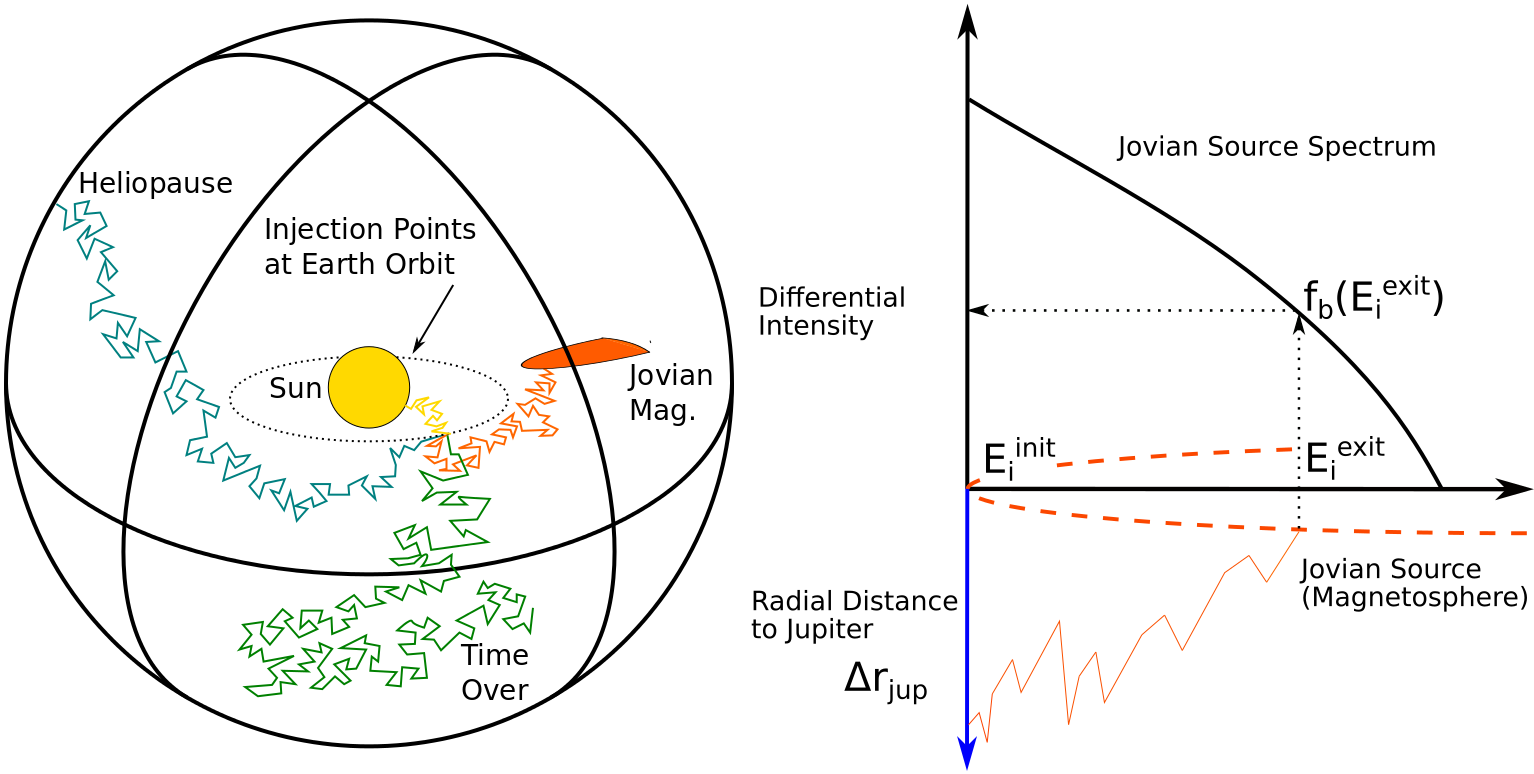}
    \caption{
    The simulation setup according to \cite{vogt2019}. On the right side a visualisation is shown of how the Jovian source spectrum determines the resulting differential intensity. The sketch thereby is to be read from bottom to top. The pseudo-particle is injected with an initial energy $E_i^{init}$ at a certain distance from the Jovian source. When the pseudo-particle's phase-space trajectory hits the source (the dashed representation of the Jovian magnetosphere) with an exit energy $E_i^{exit}$ it is convoluted with the source spectrum as the boundary condition. Subsequently the differential intensity is calculated as $\sum_i f_b(E_i^{exit})$. The dashed representation of the Jovian magnetosphere only serves an illustrative purpose.}
    \label{fig:simulation}
\end{figure*}

Knowing the shape and strength of the source as accurately as possible is important in order to normalize simulation results. In case of solving the \ac{TPE} by \cite{parker1965} by the means of \acp{SDE} the relation appears as follows: instead of solving the \ac{TPE} itself as a differential equation, a corresponding set of 
Langevin
equations is solved as described, for instance, by \cite{strauss2011}: 
\begin{equation}
\label{eqn:sde_derived}
\begin{split}
dr =&\left[\frac{1}{r^2}\frac{\partial}{\partial r}(r^2\kappa_{rr})+\frac{1}{r\sin\theta}\frac{\partial\kappa_{r\phi}}{\partial\phi}-u_{SW}\right]ds  \\
&+\sqrt{2\kappa_{rr}-\frac{2\kappa^2_{r\phi}}{\kappa_{\phi\phi}}}d\omega_r + \frac{\sqrt{2\kappa_{r\phi}}}{\kappa_{\phi\phi}}d\omega_{\phi}\\
d\theta=&\left[\frac{1}{r^2\sin\theta}\frac{\partial}{\partial\theta}(\sin\theta\kappa_{\theta\theta})\right]ds + \frac{\sqrt{2\kappa_{\theta\theta}}}{r} d\omega_{\theta}\\
d\phi=&\left[\frac{1}{r^2\sin^2\theta}\frac{\partial\kappa_{\phi\phi}}{\partial\phi}+\frac{1}{r^2\sin\theta}\frac{\partial}{\partial r}(r\kappa_{r\phi})\right]ds + \frac{\sqrt{2\kappa_{\phi\phi}}}{r\sin\theta}d\omega_{\phi}\\
dE=&\left[\frac{1}{3r^2}\frac{\partial}{\partial r}(r^2 u_{SW})\Gamma E\right]ds\mbox{,}
\end{split}
\end{equation}
with $\Gamma=(E+2E_0)/(E+E_0)$,  $\kappa_{rr/r\phi/\phi\phi/\theta\theta}$ being the entries of the diffusion tensor $\hat{\kappa}$ in spherical coordinates aligned to the \ac{HMF} as derived e.g. by \cite{Burger2000} and $ds$ the time-backward time increment. 
The adiabatic energy changes are shown as only depending on the radial position due to the assumption made here of a constant, radial Solar wind speed, a reasonable assumption in the region and on the timescales of interest to the first part of this study \cite[see, e.g.,][]{Kohnlein96}.
In order to solve Eq.~(\ref{eqn:sde_derived})
an iterative Euler-Maruyama scheme is used to solve these equations, which produces a set of random walk type solutions. This is due to the way in which the \ac{SDE} approach treats the stochastic nature of diffusion 
by implementing a
Wiener process $d\omega=\zeta\sqrt{dt}$, 
with $\zeta$ being 
a
stochastic element as a vector of Gaussian distributed random numbers. The solutions can be described as point-like solutions of the time-dependent distribution function $f(\vec{r},t)$ covering the whole phase space as described by the \ac{TPE}. 
As illustrated in Fig.~\ref{fig:simulation}, the temporal evolution of the solutions form trajectories through the phase space. In the case of a time-backward approac,h as utilised for this study \citep[see e.g.][]{Kopp2012,Strauss2017}, and depicted by Fig.~\ref{fig:simulation}, the injection point can be understood as the observational point and the phase-space trajectories follow a random walk to three possible sources, which are implemented as boundary conditions. These, according to \citep{Dunzlaff2015}, are: 
\begin{itemize}
\item 
The Sun, indicated in yellow (such trajectories are neglected in this study due to the lack of a significant Solar component in high-keV and low-Mev energy range of electrons)
    \item 
    The Jovian magnetosphere, indicated in orange (source spectrum according to \cite{vogt2018})
    \item 
    The Heliopause, indicated in blue (the local interstellar spectrum of galactic cosmic-ray electrons)
\end{itemize}
Due to the time-backward approach the amount of particles which are injected at the observational points (and later evaluated) can be held constant, whereas the percentage which reaches the different boundaries varies depending on the propagation conditions.

Usually 
the
point-like solutions 
of the \acp{SDE}
in space and time are interpreted as a set of pseudo-particles. 
Thus, the phase-space trajectories, as illustrated in Fig.~\ref{fig:simulation}, could be understood to be describing the propagation of particle samples. But
this interpretation bears the problem that the solutions of the \acp{SDE} are inherently of a mathematical nature and 
could
represent a varying number of physical particles  \citep[see][amongst others]{Kopp2012, Strauss2017,vogt2019}. Therefore 
if
we 
refer to the temporal evolution of the solutions in the following 
we will
address them plainly as phase-space trajectories. 


Since the phase space trajectories are mathematical solutions they are (and in contrast to a physical interpretation) equally probable. In order to be comparable with spacecraft data therefore a large number of phase space trajectories has to be sampled 
and weighed according to their physical significance in order
to approximate a spatial solution 
\cite[see, e.g.,][]{MolotoEA19}
. Thereby the corresponding source spectrum serves as a boundary weight  \citep[see, e. g.,][]{strauss2011,Kopp2012} in order to determine the contribution to the differential intensity as a physical quantity (or the physical likelihood of the phase space trajectory, respectively). 
As illustrated on the right side of Fig.~\ref{fig:simulation} each radial step results in adiabatic energy losses (or gains in the time-backward approach) as obtained by 
the implementation of 
the modelling approach according to Eq.~(\ref{eqn:sde_derived}) until the boundary is reached. 
Subsequently,
the trajectories are weighted 
by
the source spectrum 
according to their
exit energies $E^{exit}$ and the number of phase-space trajectories $N$. 
Physically, this refers to the fact that the
strength of the source indicates how likely a particle measured at the observational point with an initial energy $E^{init}$ has left the source with $E^{exit}$. This means, as sketched by Fig.~\ref{fig:simulation}, that the phase-space trajectories effectively are weighted by the differential intensities at the source corresponding to their adiabatic energy gain 
which defines their energy at the source
.
In case of the Jovian source spectrum $j_{jov}(E)$ this leads to the expression 
\begin{equation}
\label{eqn:diff_intensities}
j_{jov}=\frac{\sum^N_{i=1}j_{jov}(E^{exit}_i)}{N}
\end{equation}
The distribution function $f$ is the solution of the \ac{TPE}, and is related to the differential intensity by $j=P^{2}f$, where $P=pc/q$ is the rigidity of the electrons corresponding to the momentum  $p$ and the charge $q$, as well as to the speed of light $c$ \cite[see, e.g.,][]{Moraal2013}.

\subsection{The numerical setup}
\label{ssec:setup}

\begin{figure*}
        \centering
        \begin{subfigure}[b]{0.49\textwidth}  
    \centering
    \includegraphics[width=\textwidth]{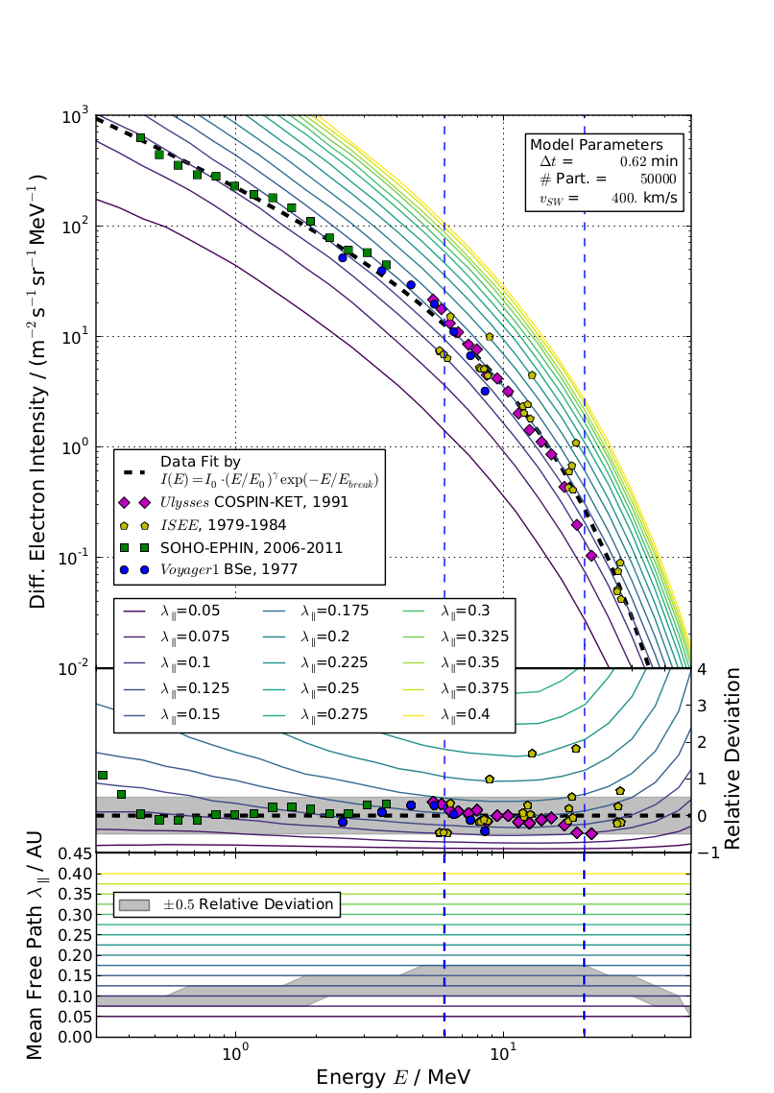}
    \caption{}
    \label{fig:spectrum_mfp_par}
        \end{subfigure} 
        \hfill        
        \begin{subfigure}[b]{0.49\textwidth}
    \centering
    \includegraphics[width=\textwidth]{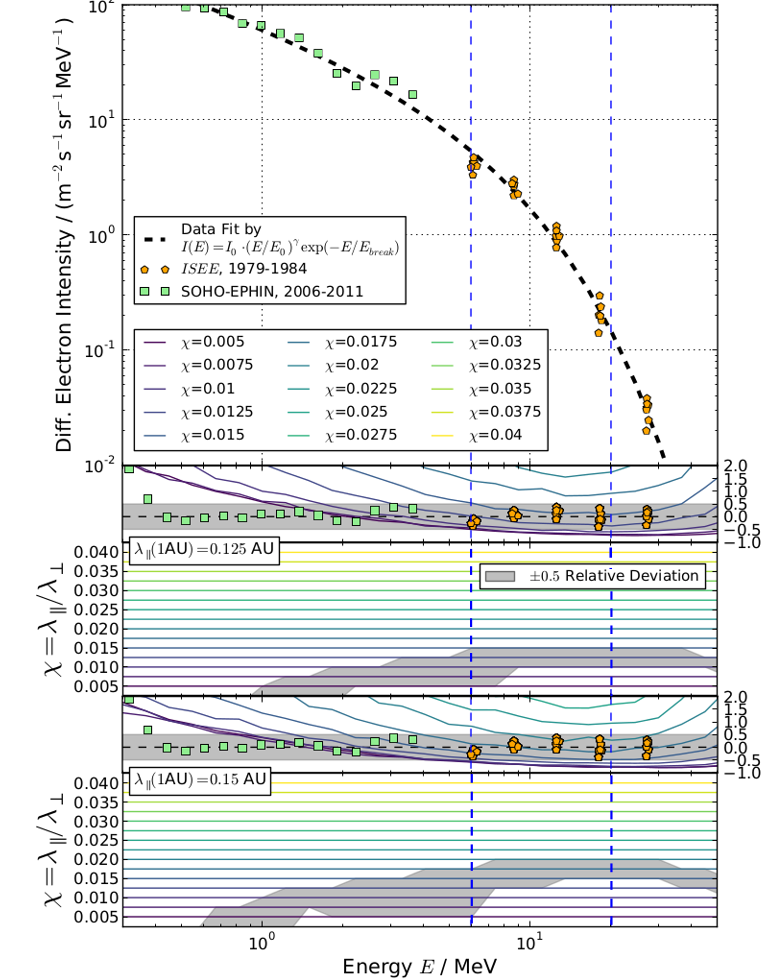}
    \caption{}
    \label{fig:spectrum_mfp_perp}
        \end{subfigure}
        \caption{
        Simulated Jovian electron spectra, similar to the Figures by \cite{vogt2019} but covering the whole energy range of the Jovian source where the energy range previously investigated by \cite{vogt2019} is indicated by the two dashed blue lines. The left panel shows on top the Earth orbit spectral data in case of good connection to the source alongside with a data fit (dashed black line). The simulation results for different values of $\lambda_{\parallel}=[0.05,0.4]~$AU are color coded. The middle panel of Fig.~\ref{fig:spectrum_mfp_par} shows the corresponding relative deviation to the source spectrum for both data and simulation results, with the area of $\pm0.5$ deviation indicated in grey. The bottom panel shows the resulting energy dependence of the parallel mean free path. The right panel displays the corresponding results for different values of $\chi=[0,0005,0.045]$ in case of bad connection. Again, the top panels show the Earth orbit spectral data alongside a data fit indicated in dashed black. The relative deviations and resulting energy dependencies are shown for $\lambda_{\parallel}=0.125~$AU (second and third panel) and $\lambda_{\parallel}=0.15~$AU (fourth and fifth panel).}
    \label{fig:mean_free_paths_data}   
\end{figure*}

In an extension of the work presented in \cite{vogt2019}, the same \ac{SDE} code as discussed by \cite{Dunzlaff2015} is utilized in order to perform the simulations. Since it is written in CUDA in order to optimize the performance time, several simplifications haven been made due to the limited internal memory on \acp{GPU}. The solar wind velocity is chosen to be directed radially outward and constant throughout the simulation setup. Following the deeper discussion on parameter setting by \cite{vogt2019}, the value for this quantity is set to $u_{SW}=400~$km/s, in agreement with solar minimum conditions. The corresponding \ac{HMF} is assumed to be a Parker field. The \ac{HMF} geometry enters into the transformation of the diffusion tensor, done here according to the approach of \cite{Burger2000}. 
Thereby it is important to note that the \ac{HMF} is only considered in this implicit way as parameters determining the geometry and values of the diffusion tensor and not implemented into the simulation setup itself as done by comparable finite difference scheme or MHD simulation setups. Furthermore the geometry of the \ac{HMF} is assumed as static throughout the simulation both as a whole and as well as regarding the individual field lines, effectively neglecting the Solar rotation. This approach is standard in Jovian transport studies and generally in cosmic ray modulation studies \cite[see, e.g.,][]{KotaJokipii83,Ferreira2002,fer03,Burger2008,Potgieter2017}. The effects of these simplifications on Jovian electron transport simulations are discussed by \cite{vogt2019} and in more detail in Sec.~\ref{sec:corotation}.

The mean free paths determining the diffusion tensor are included and are based on the simplified, energy-independent approach applied successfully by \citep[e. g.][amongst others]{strauss2011,strauss2011b}:  $\lambda_{\parallel}$ is normalised  at $r_0=1$~AU, such that
\begin{equation}
\label{eqn:lambda_par}
\lambda_{\parallel}(r)=\frac{\lambda_0}{2}\left(1+\frac{r}{r_0}\right)\mbox{.}
\end{equation}
An energy dependence, however, does enter into the diffusion coefficient by way of its standard definition, viz. $\kappa=v\lambda/3$ \cite[see, e.g.,][]{shalchibook}, 
with $v$ being the particle speed
. Again following previous studies 
\cite[e.g.][]{Ferreira2001,Ferreira2001b}
, the perpendicular diffusion coefficient is assumed to be proportional to the parallel diffusion coefficient by a constant factor $\chi$ such that
\begin{equation}
\kappa_{\perp}(r)=\chi\kappa_{\parallel}(r)\mathrm{.}
\end{equation}
For the purposes of the present study, these simplified transport parameters 
serve to qualitatively emulate the behaviour in the inner heliosphere of more theoretically-motivated expressions for the same employed in \textit{ab initio} particle transport studies \citep[see, e.g.,][]{EB2013,Engelbrecht2019}. For instance, the energy-independence of $\lambda_{\parallel}$ over the range of (low) energies considered in this study reflects the behaviour of electron parallel mean free paths derived from quasilinear theory \citep[see, e.g.,][]{ts2003,EB2010,DempersEngelbrecht20}, as well as what is expected from observations \citep[e.g.][]{Bieber1994}. The energy-independent perpendicular mean free path is chosen to reflect the energy dependence expected from observational \citep[e.g.][]{Palmer1982} and modulation studies \citep[e.g.][]{Burger2000} at the energies relevant to this study, and does not differ much from the relatively weak energy dependences of perpendicular mean free paths derived from theory \citep[see, e.g.,][]{shalchibook,Burger2008,EB2015,Engelbrecht2019}.

Tab.~\ref{tab:simulation_parameters} shows the choice of parameters used within this framework as motivated and tested by \cite{vogt2019}.
In order to do so, those authors performed a set of parameter studies varying the 'computational' parameters. The set of values listed in Tab.\ref{tab:simulation_parameters} was found to be a trustworthy compromise between the need to keep the simulation time low and to assure that the results are converging. Also the radius of the model setup $R_{HP}$ is chosen specifically because a smaller setup can influence the simulation results, as demonstrated by that study. The parallel mean free path $\lambda_{\parallel}$ was determined by simulating the Jovian electron spectrum at Earth orbit during times of good magnetic connection at which the particle transport can be assumed to be dominated by parallel diffusion. 
Fig.~\ref{fig:mean_free_paths_data} shows the results of this study over an extended energy range, covering the whole range of interest of the Jovian source spectrum, as opposed to the single energy considered by \cite{vogt2019}. For the case of good magnetic connection to the source, the parallel mean free path $\lambda_{\parallel}$ seems to be only slightly energy-dependent within the constraints of our modelling approach, although in contrast to the limited energy range investigated by \citep{vogt2019} $\lambda_{\parallel}=0.125~$AU appears to be more in agreement with the simulation results covering the extended energy range.
Therefore
, the value listed in Tab.~\ref{tab:simulation_parameters} appears to produce results that are most in agreement with spacecraft data at Earth orbit during comparable conditions. Likewise $\chi$ was determined by applying the result for $\lambda_{\parallel}$ and comparing the results of simulation runs with different values of $\chi$ to Jovian electron spectra obtained during times of poor magnetic connection 
as shown in Fig.~\ref{fig:spectrum_mfp_perp}, where perpendicular diffusion is expected to be the predominant mechanism governing the transport of these electrons. As can be seen for both cases of $\lambda_{\parallel}=0.125~$AU (second and third panel) and $\lambda_{\parallel}=0.15~$AU (fourth and fifth panel) 
the energy dependence of $\chi$ is more prominent. It is worth noting that the decrease of $\chi$ is located in the keV-range. As discussed by \cite{vogt2018}, the Jovian source spectrum below the MeV range can only be determined by utilizing $\textit{Pioneer 10}$ flyby data by \cite{Teegarden1974}, which has to be re-normalised, and \textit{SOHO} data by \cite{kuehl2013} at Earth orbit, which shows a variation in its steepness below $1~$MeV. These two uncertain influences on the source spectrum could translate into uncertainty of the behaviour of $\chi$ at these energies. 
In order to maximise the energy range with a physically reasonable choice, $\chi=0.0125$ was used for the simulation within this study as listed alongside the other parameters in Tab.~\ref{tab:simulation_parameters}. 
The lack of energy dependence assumed in the simulation setup for the perpendicular mean free path $\lambda_{\perp}$ is not expected to cause significant effects. As shown by \cite{vogt2019} the range of exit energies which are significant in order to obtain the corresponding differential intensities is much smaller than the ones relevant for the energy dependence of $\chi$ in Fig.~\ref{fig:spectrum_mfp_perp}.

\begin{table}
\centering
\caption{The computational and physical parameters used for the  code (when not stated otherwise).} 
\begin{tabular}{l l}
\hline\hline
\multicolumn{2}{l}{Computational Parameters}\\ \hline
\# Trajectories & 50000 \\
$\Delta t$ & 0.0001\\
$T_{End}$ & 800\\\hline
\multicolumn{2}{l}{Physical Parameters}\\ \hline
$R_{HP}$ & $120~$AU \\
$u_{SW}$ & $400~$km/s\\
$\lambda_{\parallel}(1~$AU$)$ & $0.125~$AU \\
$\chi=\lambda_{\perp}/\lambda_{\parallel}$ & $0.0125$\\
$E^{init}$ & $6~$MeV\\
\end{tabular} 
\\[10pt]
\label{tab:simulation_parameters}
\end{table}


\subsection{The influence of boundary conditions}
\label{ssec:boundary_conditons}
\begin{figure*}
        \centering
        \begin{subfigure}[b]{0.49\textwidth}  
            \centering 
            \includegraphics[width=\textwidth]{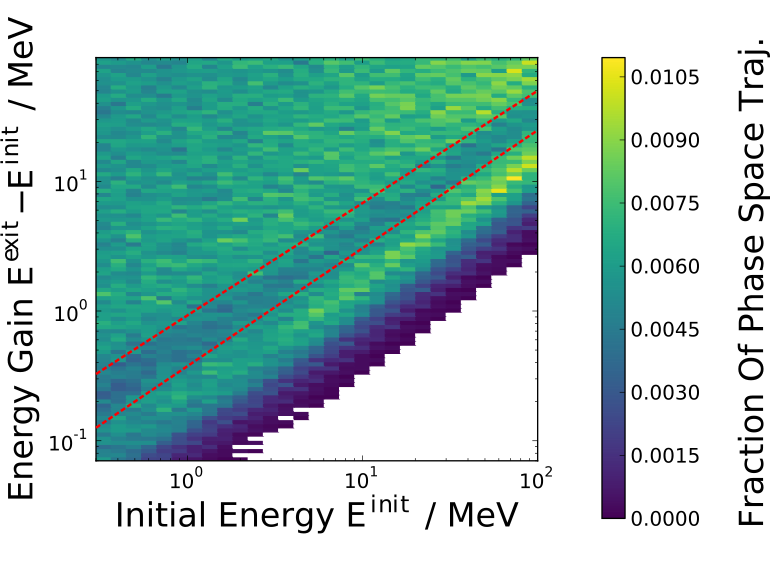}
            \caption{}    
            \label{fig:influence_spectrum_vs_energy_traj_best}
        \end{subfigure} 
        \hfill        
        \begin{subfigure}[b]{0.49\textwidth}
            \centering
            \includegraphics[width=\textwidth]{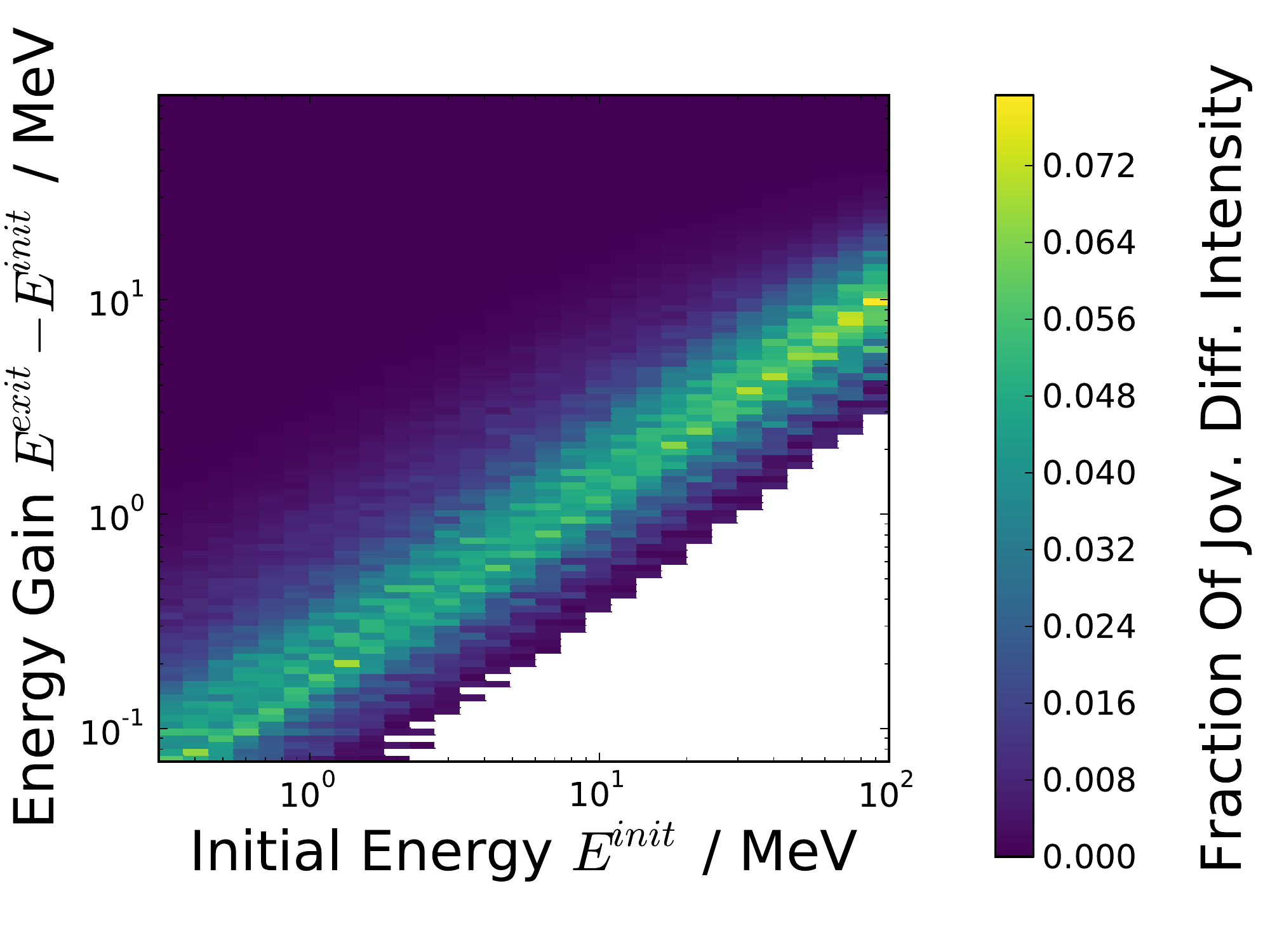}
            \caption{}    
            \label{fig:influence_spectrum_vs_energy_flux_best}
        \end{subfigure}
                        \vskip\baselineskip
        \centering
        \begin{subfigure}[b]{0.49\textwidth}  
            \centering 
            \includegraphics[width=\textwidth]{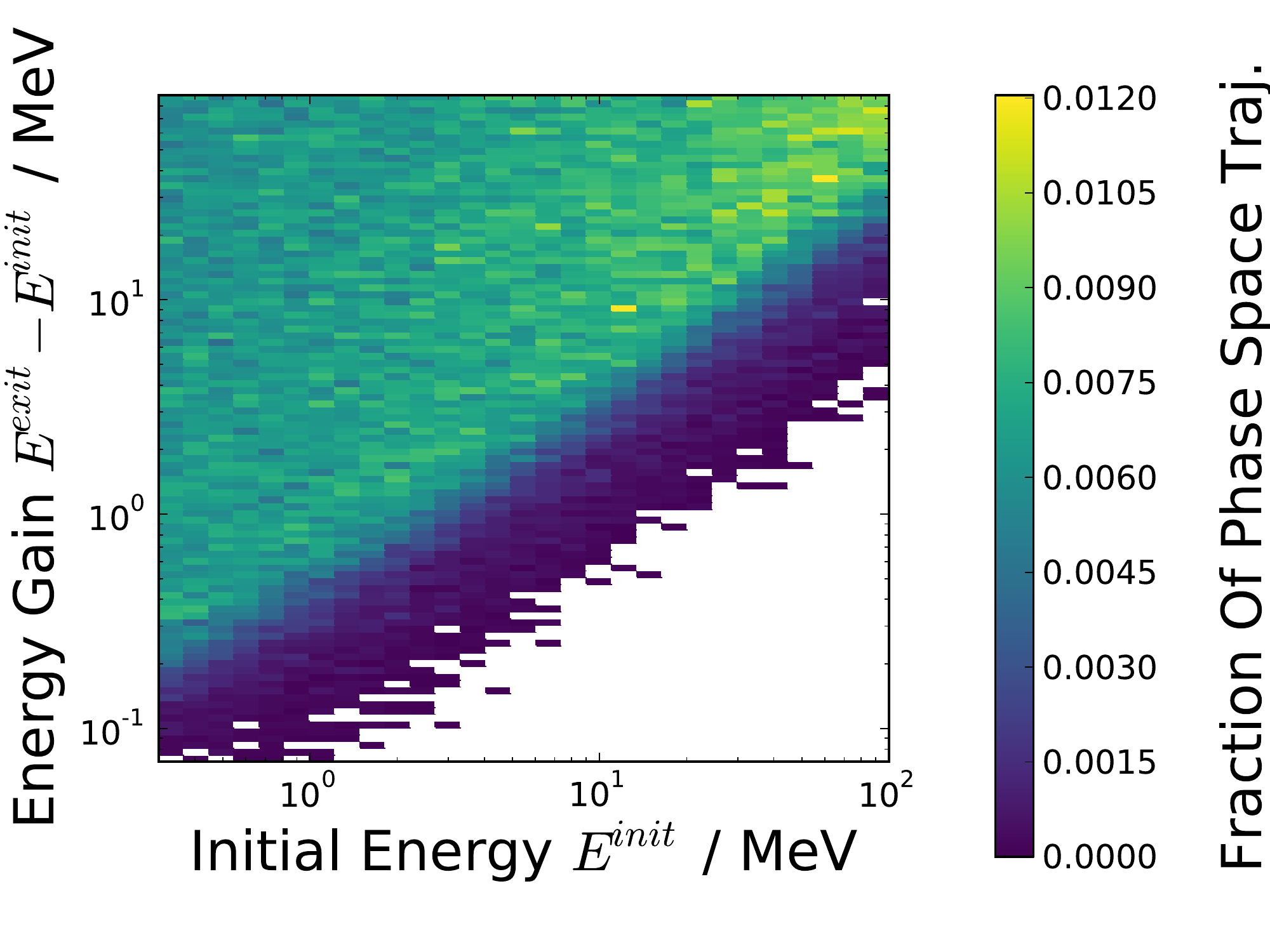}
            \caption{}    
            \label{fig:influence_spectrum_vs_energy_traj_worst}
        \end{subfigure} 
        \hfill        
        \begin{subfigure}[b]{0.49\textwidth}
            \centering
            \includegraphics[width=\textwidth]{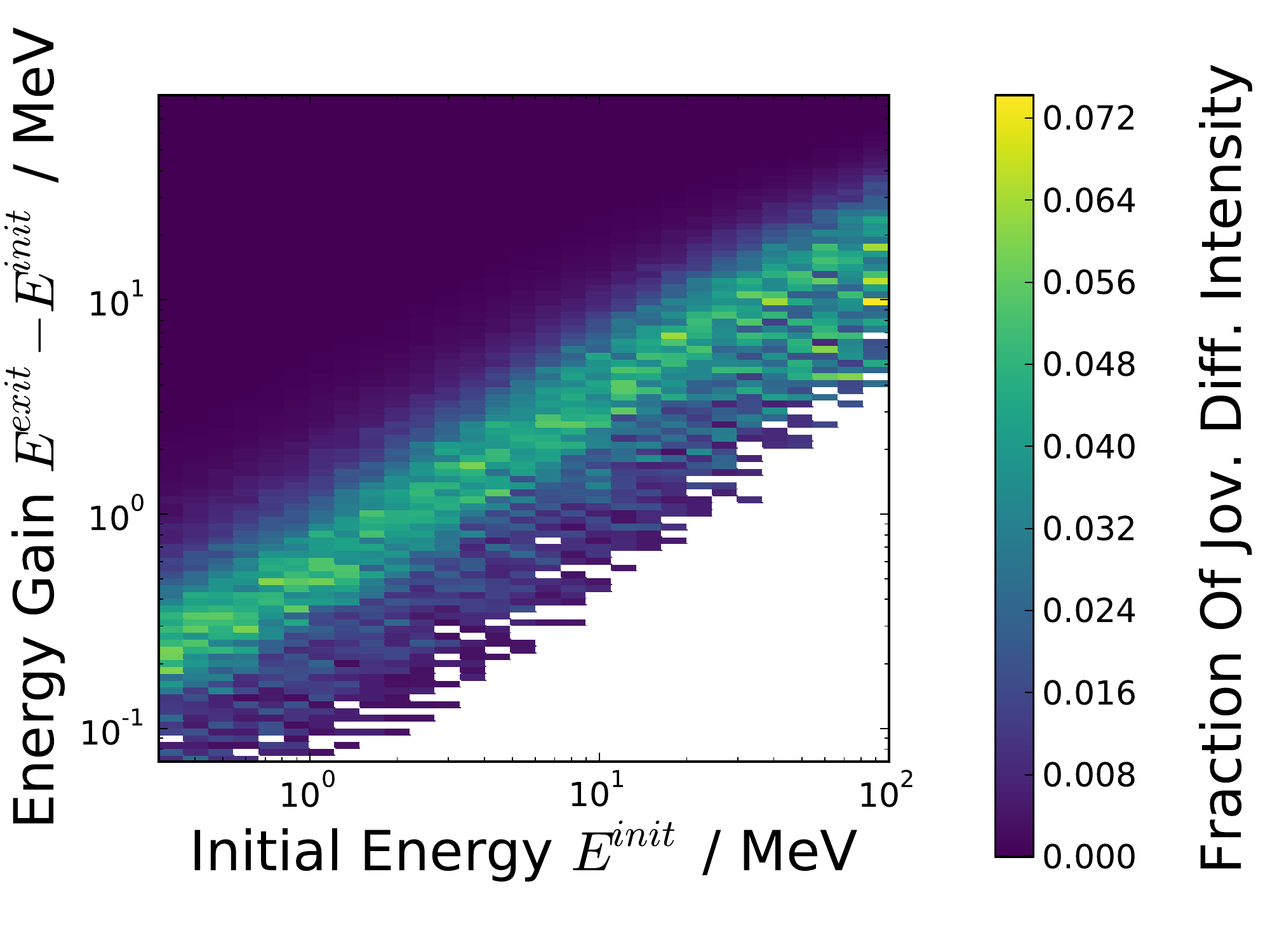}
            \caption{}    
            \label{fig:influence_spectrum_vs_energy_flux_worst}
        \end{subfigure}        
                         \vskip\baselineskip
        \centering
        \begin{subfigure}[b]{0.49\textwidth}  
            \centering 
            \includegraphics[width=\textwidth]{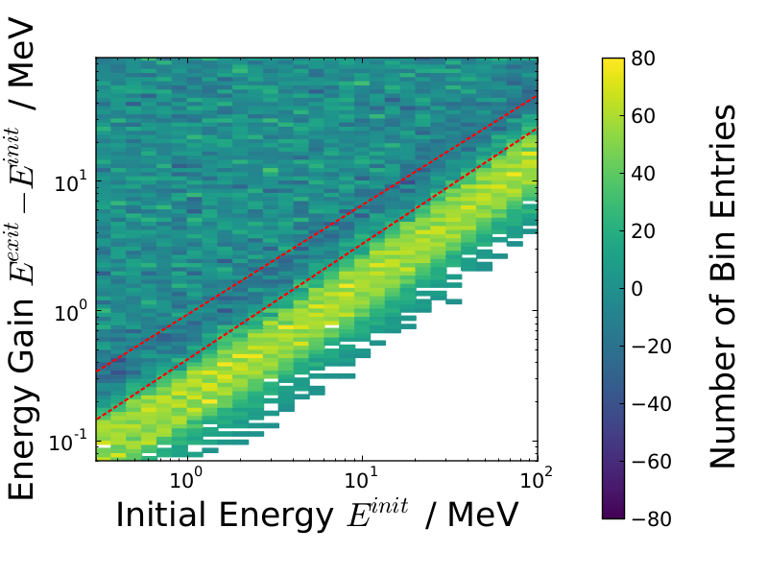}
            \caption{}    
            \label{fig:influence_spectrum_vs_energy_traj_compare}
        \end{subfigure} 
        \hfill        
        \begin{subfigure}[b]{0.49\textwidth}
            \centering
            \includegraphics[width=\textwidth]{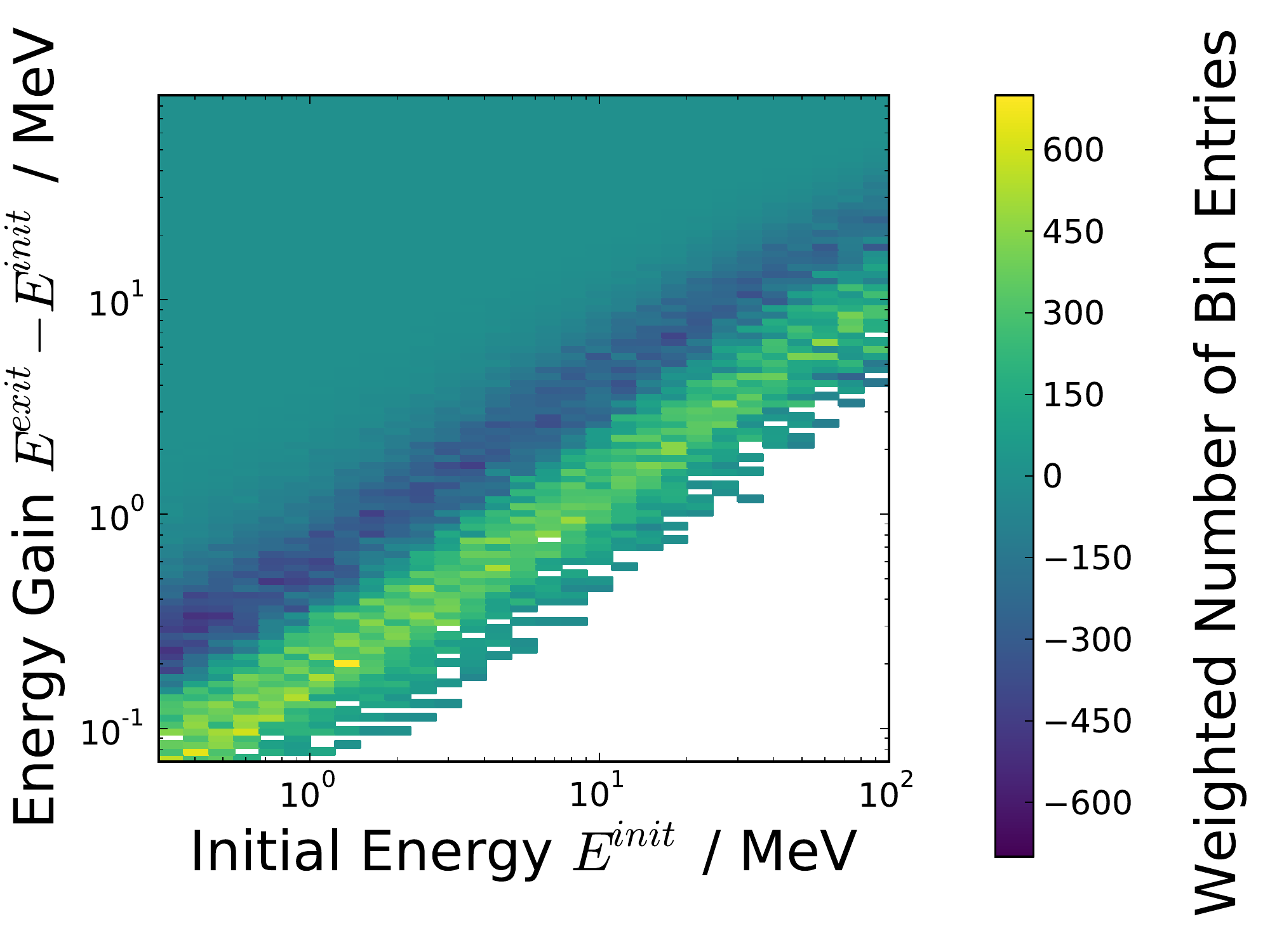}
            \caption{}    
            \label{fig:influence_spectrum_vs_energy_flux_compare}
        \end{subfigure}        
    \caption{
    Binned distributions of exit energies $E_i^{init}$ over the whole energy range dominated by Jovian electrons. Whereas the right panels show these distributions weighted by their contribution according to Eq.~(\ref{eqn:diff_intensities}), the left side displays the unweighted results of the simulation. The upper most panels Figs.~\ref{fig:influence_spectrum_vs_energy_traj_best} and \ref{fig:influence_spectrum_vs_energy_flux_best} show the case of good magnetic connection to the source, whereas Figs.~\ref{fig:influence_spectrum_vs_energy_traj_worst} and \ref{fig:influence_spectrum_vs_energy_flux_worst} show the case of bad magnetic connection. The differences between those distributions are depicted by the two lower panels, Figs.~\ref{fig:influence_spectrum_vs_energy_traj_compare} and \ref{fig:influence_spectrum_vs_energy_flux_compare}.}
    \label{fig:influence_spectrum_vs_energy_flux}
\end{figure*}

The effect of applying Eq.~(\ref{eqn:diff_intensities}) is shown by Fig.~\ref{fig:influence_spectrum_vs_energy_flux}. The panels on the left display the binned distribution of phase-space trajectories according to their exit energies for good (top panel) and poor magnetic connection to the source (middle panel) as well as the difference between them, assuming a Parker field 
geometry with a constant radial Solar wind speed
. On the right side the corresponding distributions are shown if Eq.~(\ref{eqn:diff_intensities}) is applied and the phase-space trajectories are therefore weighted according to their physical significance. In both cases the distributions are normalized to one. In order to interpret the results, 
particularly
the relation between the distributions of the trajectories and the differential intensities, Fig.~\ref{fig:influence_spectrum_vs_energy_traj_best} displays an important feature: As marked by two dashed red lines, a sink in the intensity can be spotted throughout the whole energy range. It is important to note that it is directly located above the highest intensities shown by the phase-space trajectories, which are associated with very low energy gains. The comparison with Fig.~\ref{fig:influence_spectrum_vs_energy_traj_worst} shows that the intensity sink in Fig.~\ref{fig:influence_spectrum_vs_energy_traj_best} appears where the phase-space distributions in the case of poor magnetic connection to the source become significant. This shift of the maximum of the distributions according to the magnetic connection to the source is also visible by Figs.~\ref{fig:influence_spectrum_vs_energy_flux_best} and \ref{fig:influence_spectrum_vs_energy_flux_worst} for the weighted contribution to the resulting differential intensities.

As noted above, both Figures display the differences between the distributions, Fig.~\ref{fig:influence_spectrum_vs_energy_traj_compare} the number of trajectories in each bin and Fig.~\ref{fig:influence_spectrum_vs_energy_flux_compare} the sum of the weight of these trajectories according to the source spectrum. For both Figures the color coding is defined symmetrically around zero. This highlights the important fact that despite some negative 
bins in the difference distributions which are mostly located within the area of the intensity sink, there is either an equal amount of or more bin entries (and therefore phase-space trajectories attributed to Jovian electrons) for the case of good magnetic connection to the Jovian source. Whereas this in itself is not an unexpected result (see Sec.~\ref{ssec:SDE_modelling}), it is remarkable that the distinct maximum of this difference distribution is located below the intensity sink in Fig.~\ref{fig:influence_spectrum_vs_energy_traj_best}, which is indicated by two dashed red lines. Therefore the maximum below the intensity sink in Fig.~\ref{fig:influence_spectrum_vs_energy_traj_best} (which is identical with the population of phase-space trajectories responsible for almost the entire contribution to the resulting differential intensity according to Fig.~\ref{fig:influence_spectrum_vs_energy_flux_best}) has to be attributed to parallel diffusion-dominated transport. Since propagation parallel to the nominal Parker spiral is the most effective mechanism for charged particle transport, it can be expected to be related to both shorter simulation times (i.e. amounts of random walk steps to perform the corresponding phase-space trajectory) as well as lower energy gains. That explains why this population of trajectories is absent in Fig.~\ref{fig:influence_spectrum_vs_energy_traj_worst}, showing the phase-space trajectory distribution in case of poor magnetic connection to the source.

We therefore interpret the simulation results as displayed in Fig.~\ref{fig:influence_spectrum_vs_energy_flux} as follows: Since the left upper triangle of the panels showing the phase-space trajectory distributions seems not to vary with respect to the degree of magnetic connection, this population can be interpreted as a kind of diffusive background. By this we aim to describe the fact that these trajectories appear to represent a population which exists, in a sense, independently of the physical propagation conditions pertaining to magnetic connectivity. This interpretation is further supported by the fact that, according to the corresponding panels on the right side of Fig.~\ref{fig:influence_spectrum_vs_energy_flux}, this population has almost no physical significance with respect to the resulting differential intensities. Therefore it can be 
shown \citep[see][]{vogt2019}
that energy gains of more than half an order of magnitude most likely correspond to phase-space trajectories of low physical probability. With respect to Figs.~\ref{fig:influence_spectrum_vs_energy_traj_compare} and \ref{fig:influence_spectrum_vs_energy_flux_compare} we can conclude that the (physical) variation of intensities 
corresponding to changing magnetic connections between the observer and the source
is caused by the (mathematical) variation of the maximum of the distribution of the phase-space trajectories below the intensity sink. 

\section{Modelling of residence times }
\label{sec:residence_times}

Within the mathematical framework of the \cite{parker1965} \ac{TPE}, the residence time corresponding to the evolution of the probability density of the particles $\rho$ is defined as the expectation value of the time according to
\begin{equation}
\label{eqn:expectation_value_time}
    \tau = \frac{\int \rho (\vec{x}, t) t dt}{\int \rho (\vec{x},t) dt}.
\end{equation}
As \cite{vogt2019} pointed out, this approach cannot be transformed into \ac{SDE} modelling by simply averaging over the pseudo-particles' exit times. Instead $\rho$ is proposed to be the distribution of particle density as 
\begin{equation}
    \rho (\vec{x},t) = \frac{f(\vec{x}^{exit},t)}{f_0(\vec{x})}
\end{equation}
and normalized by the total phase-space density
\begin{equation}
    f_0 (\vec{x}) = \int f(\vec{x}^{exit},t) dt.
\end{equation}
Thereby $f_0(\vec{x})$ cancels out in Eq.~(\ref{eqn:expectation_value_time}). Applied to the discrete case of phase space trajectories as the output of an \ac{SDE} code,  Eq.~(\ref{eqn:expectation_value_time}) reads as 
\begin{equation}
\label{eqn:mean_prop_time}
\tau(r^0,E^0) =\frac{\sum^N_{i=1} s(E^{exit}_i)\cdot f(E^{exit}_i)}{\sum^N_{i=1}f(E^{exit}_i)},
\end{equation}
with $s$ as the exit (integration) time corresponding to the different phase space trajectories. 
For a more detailed discussion of the derivation see Sec.~3 of \cite{vogt2019}, especially Eqs.~(7) to (10).
Compared to the calculation of differential intensities by applying Eq.~\ref{eqn:diff_intensities}, this approach is self consistent, as each phase space trajectory is attributed the same significance due to it's exit energy in both cases. 
Note for the further discussion within this paper, that the exit energies are the results of the adiabatic energy changes experienced throughout the random walk which constitute the phase-space trajectories. Therefore the adiabatic energy changes define - in combination with the source spectrum as shown by Fig.~\ref{fig:simulation} - the weight by which the exit times of the individual phase-space trajectories are taken into account.
For 6 MeV Jovian electrons, \cite{vogt2019} show further that the residence times obtained via Eq.~(\ref{eqn:mean_prop_time}) are between $\sim 5 - 11$ days, depending on magnetic connectivity, and are therefore shorter by almost two orders of magnitude when compared to previous estimations \citep[see e.g.][and references therein]{strauss2011b,vogt2019}.

\subsection{Adiabatic energy changes and the role of the source spectrum}
\label{ssec:adiabatic_cooling}

\begin{figure*}
        \centering
        \begin{subfigure}[b]{0.49\textwidth}  
            \centering 
            \includegraphics[width=\textwidth]{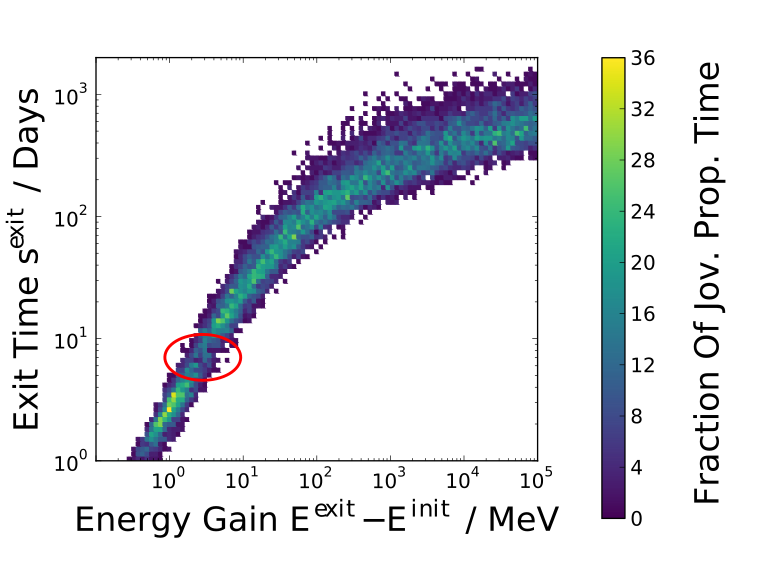}
            \caption[]%
            {}    
            \label{fig:exit_times_vs_energy_sexit}
        \end{subfigure} 
        \hfill        
        \begin{subfigure}[b]{0.49\textwidth}
            \centering
            \includegraphics[width=\textwidth]{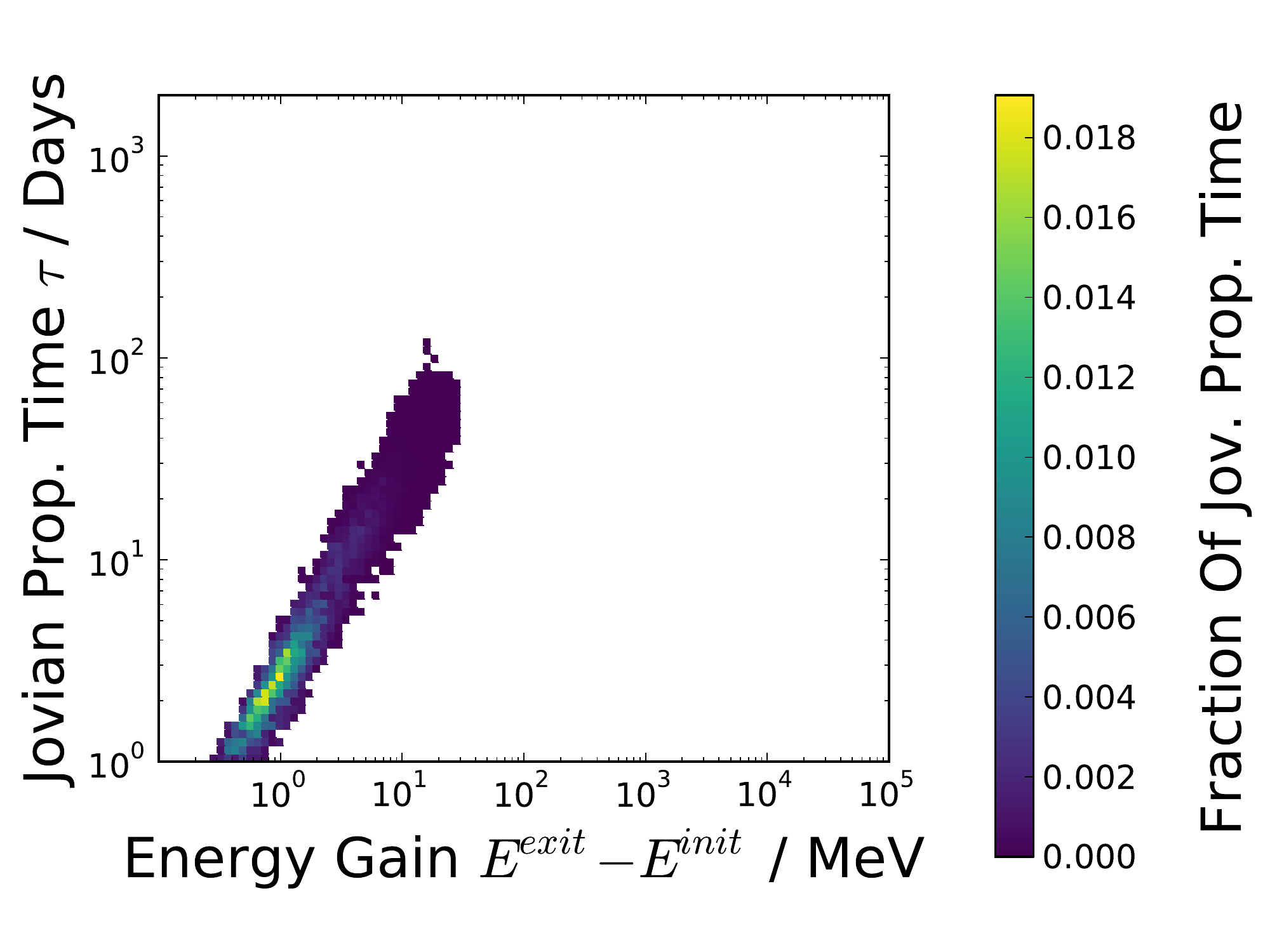}
            \caption[]%
            {}    
            \label{fig:exit_times_vs_energy_tau}
        \end{subfigure}
        \caption{The relation between the energy gain $E^{exit}_i-E^{init}_i$ (or loss, in the normal time-forward scenario) as a function of the corresponding exit time $s^{exit}_i$ (left panel) or propagation time (right panel), weighted by applying Eq.~(\ref{eqn:mean_prop_time}). The case of good magnetic connectivity is shown.}
        \label{fig:exit_times_vs_energy}
\end{figure*} 

According to Eq.~(\ref{eqn:mean_prop_time}) the contribution of each phase-space trajectory depends on its exit time $s(E^{exit}_i)$, 
more
precisely on the differential intensity $f(E^{exit}_i)$ corresponding to it within the source spectrum. We aim to investigate this relation further. 
Fig.~\ref{fig:exit_times_vs_energy} shows the relation between the energy gain and the exit times of the phase space trajectories unweighted by Eq.~(\ref{eqn:mean_prop_time}) (Fig.~\ref{fig:exit_times_vs_energy_sexit}) and weighted by their contribution to the residence time (Fig.~\ref{fig:exit_times_vs_energy_tau}) for the case of good magnetic connectivity between the Jovian source and the observer. 
As discussed above with reference to Fig.~\ref{fig:influence_spectrum_vs_energy_flux}
, the energy differences between the initial energies $E^{init}_i$ and the exit energies $E^{exit}_i$ are caused by adiabatic energy changes as 
given
by \cite{parker1965} within the \ac{TPE}. For the \acp{SDE} modelling approach as utilized here, the influence of this effect 
is included
as follows:  Whereas each subsequent iteration of the random walk increases the corresponding simulation time $s_i$ by the the value of the time increment, the energy gain accompanying each step of the phase space trajectory is a function of energy and position. Therefore phase space trajectories of the same (time) length can be related to different rates of adiabatic energy changes $E^{exit}-E^{init}$ as illustrated in the right upper part of Fig.~\ref{fig:exit_times_vs_energy_sexit}. 
As given by
\begin{equation}
\label{eqn:adiabatic_cooling}
    \frac{1}{E}\frac{dE}{dt}=-\frac{1}{3}\nabla\cdot \vec{u_{SW}}\mbox{,}
\end{equation}
 according to \cite{parker1965} and \cite{Webb1979}, the relation of adiabatic energy changes to simulation times is almost linear in the logarithmic display of Fig.~\ref{fig:exit_times_vs_energy_tau}. 
 Since Fig.~\ref{fig:exit_times_vs_energy_sexit} consists of the simulation results displayed with respect to their initial energies $E_i^{init}$ in Fig.~\ref{fig:influence_spectrum_vs_energy_flux_best}, the intensity sink discussed in Sec.~\ref{ssec:SDE_modelling} also appears in Fig.~\ref{fig:exit_times_vs_energy_sexit}. The depiction (especially with respect to Fig.~\ref{fig:exit_times_vs_energy_tau}) allows one to deduce that, according to the location of the intensity sink (red), phase-space trajectories corresponding to energy gains of more than half of an order of magnitude and a simulation time of more than eight days do not contribute significantly to the modelling results. This notion further supports the interpretation of Fig.~\ref{fig:influence_spectrum_vs_energy_flux} in Sec.~\ref{ssec:SDE_modelling} that this intensity sink segregates the physically significant population of phase-space trajectories from the (diffusive background) population which is shown in Fig.\ref{fig:exit_times_vs_energy_sexit} as a high energy gain and high exit time tail. The comparison with Fig.~\ref{fig:exit_times_vs_energy_tau} validates this, as the high energy gain and high exit time tail is shown to weighted down to effectively zero by the boundary condition, i.e. the Jovian source spectrum.  
 
 These considerations suggest that the adiabatic energy changes implemented according to Eq.~(\ref{eqn:adiabatic_cooling}) always have to be considered in order to obtain physically reliable results as long as diffusive processes influence the particle's propagation. As shown by Figs.~\ref{fig:influence_spectrum_vs_energy_flux} and \ref{fig:exit_times_vs_energy} and discussed in detail in Sec.~\ref{ssec:SDE_modelling}, the physical likeliness of (mathematically) equally possible phase-space trajectories is determined by the boundary conditions via the corresponding exit energy $E_i^{exit}$. Therefore, not considering adiabatic energy changes effectively assumes that the energy spectrum of the particle population of interest is flat, and therefore each energy is equally probable. This relation is also shown on the right side of Fig.~\ref{fig:simulation}. The following Section investigates further the significant influence of the boundary conditions (and, by implication, adiabatic energy changes) on  Jovian residence times over the whole Jovian energy range. Subsequently, these considerations and results will be applied to the analytical estimation of the \ac{GCR} residence times as suggested by, e.g., \cite{parker1965,OGallagher1975,strauss2011b}. 

\subsection{
Adiabatic
Energy dependence of 
Jovian Residence Times}
\label{ssec:energy_dep}

\begin{figure*}
        \centering
        \begin{subfigure}[b]{0.49\textwidth}  
            \centering 
            \includegraphics[width=\textwidth]{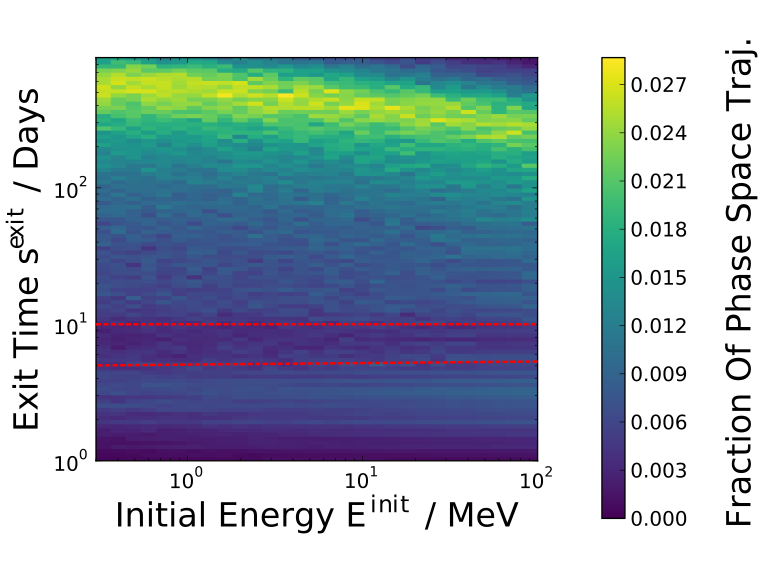}
            \caption[]%
            {
            }    
            \label{fig:influence_spectrum_vs_energy_meantime_best}
        \end{subfigure} 
        \hfill        
        \begin{subfigure}[b]{0.49\textwidth}
            \centering
            \includegraphics[width=\textwidth]{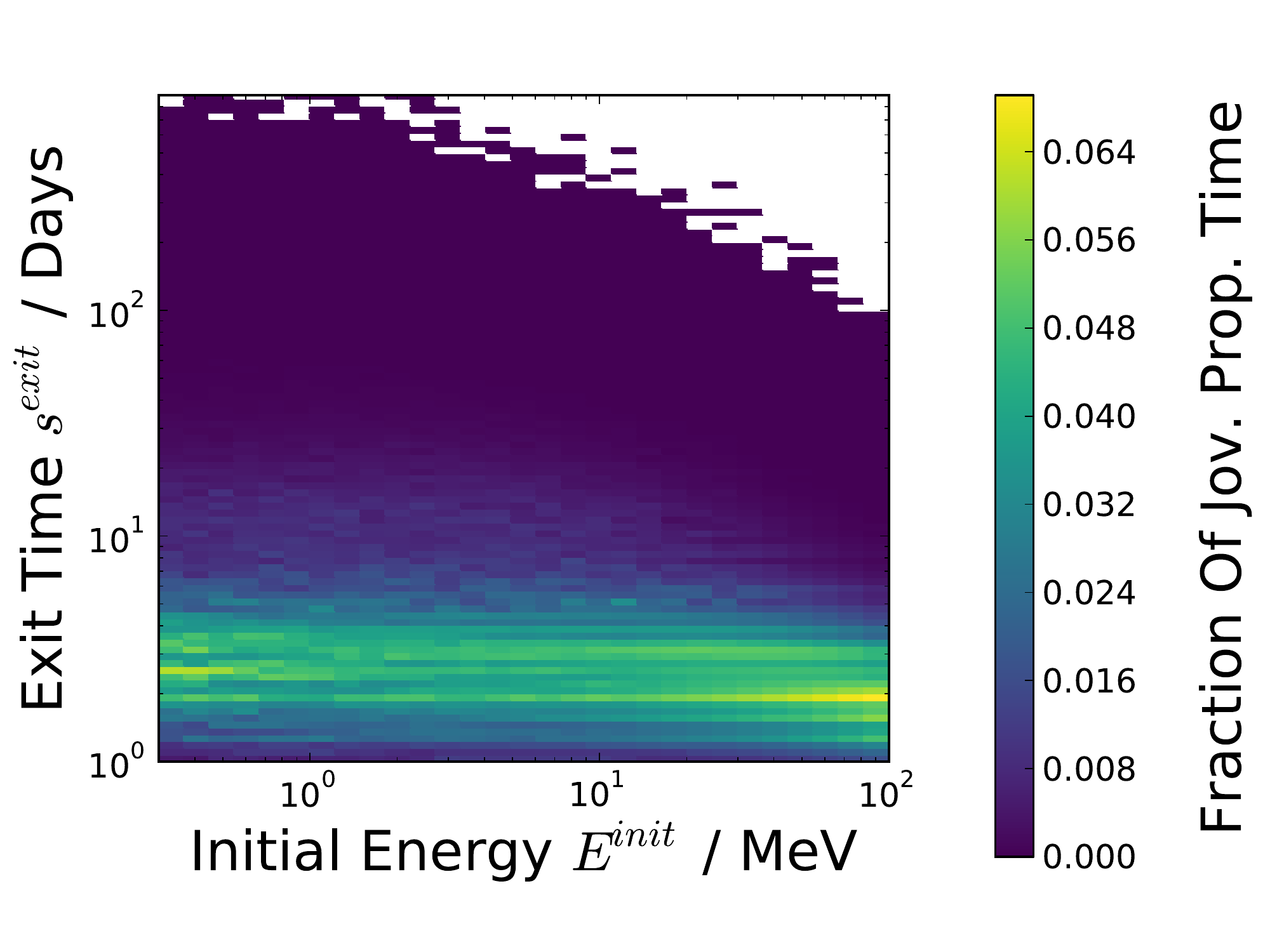}
            \caption[Network2]%
            {
            }    
            \label{fig:influence_spectrum_vs_energy_mpt_best}
        \end{subfigure}
                        \vskip\baselineskip
        \centering
        \begin{subfigure}[b]{0.49\textwidth}  
            \centering 
            \includegraphics[width=\textwidth]{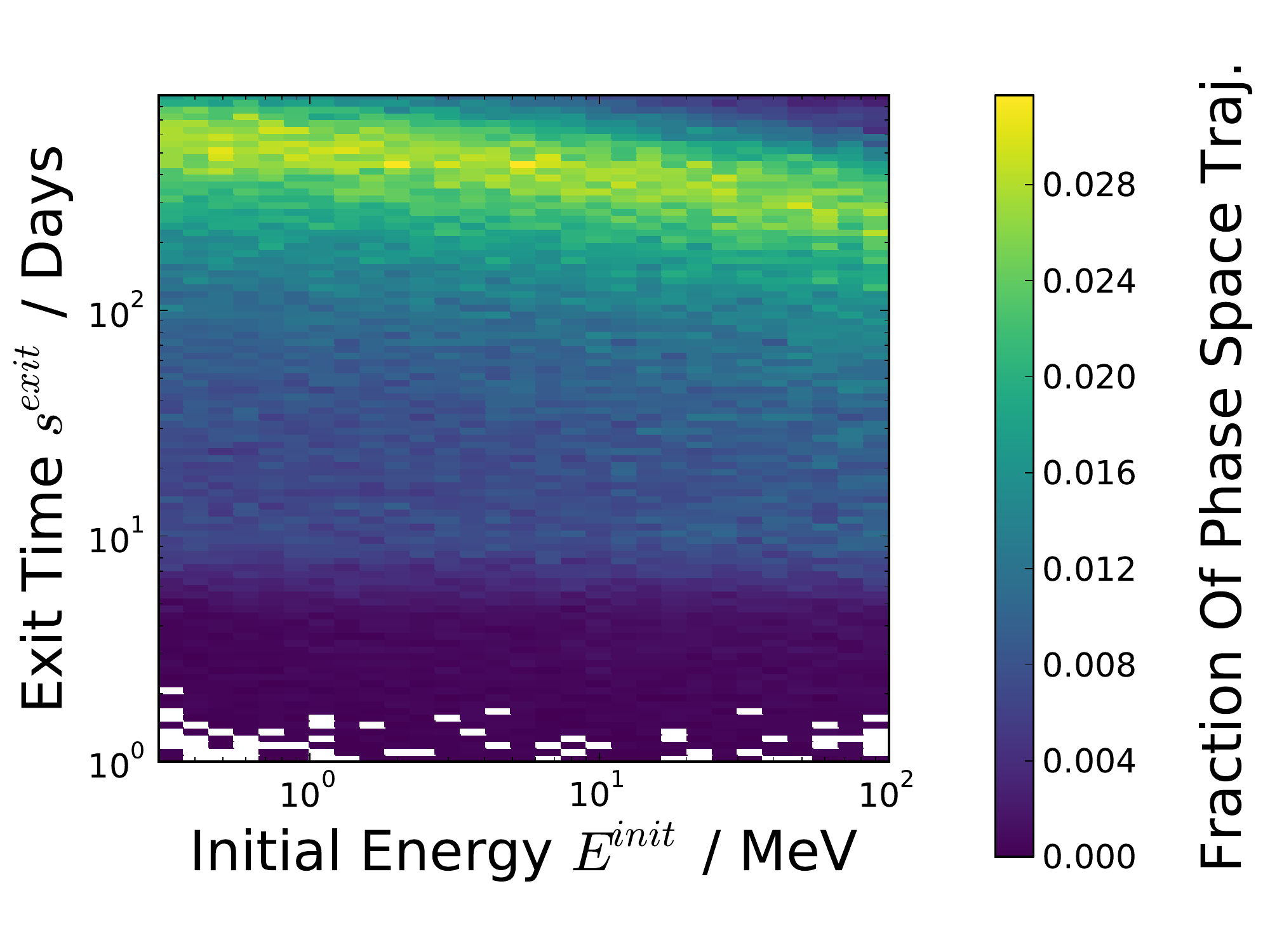}
            \caption[]%
            {
            }    
            \label{fig:influence_spectrum_vs_energy_meantime_worst}
        \end{subfigure} 
        \hfill        
        \begin{subfigure}[b]{0.49\textwidth}
            \centering
            \includegraphics[width=\textwidth]{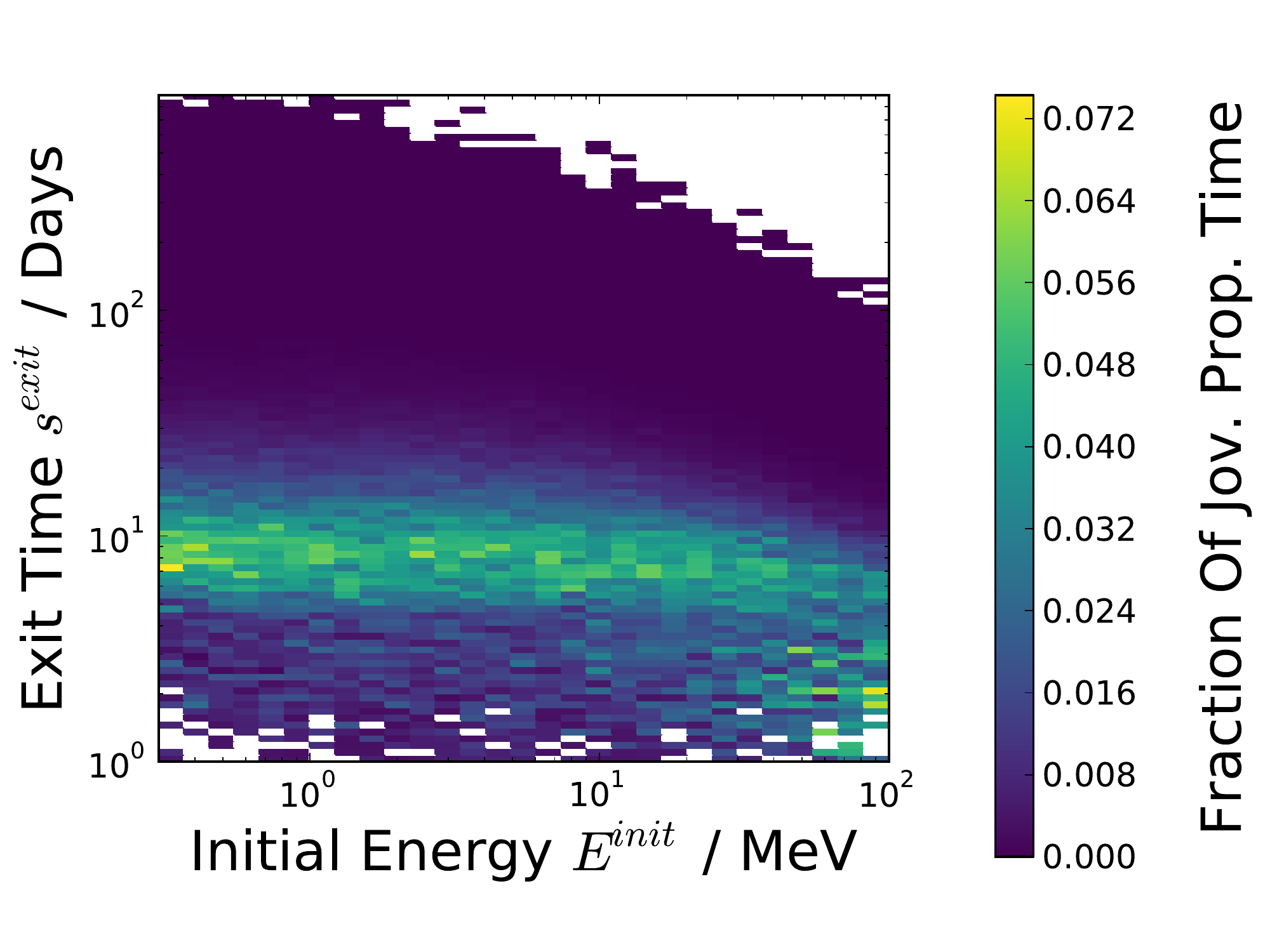}
            \caption[Network2]%
            {
            }    
            \label{fig:influence_spectrum_vs_energy_mpt_worst}
        \end{subfigure}        
                         \vskip\baselineskip
        \centering
        \begin{subfigure}[b]{0.49\textwidth}  
            \centering 
            \includegraphics[width=\textwidth]{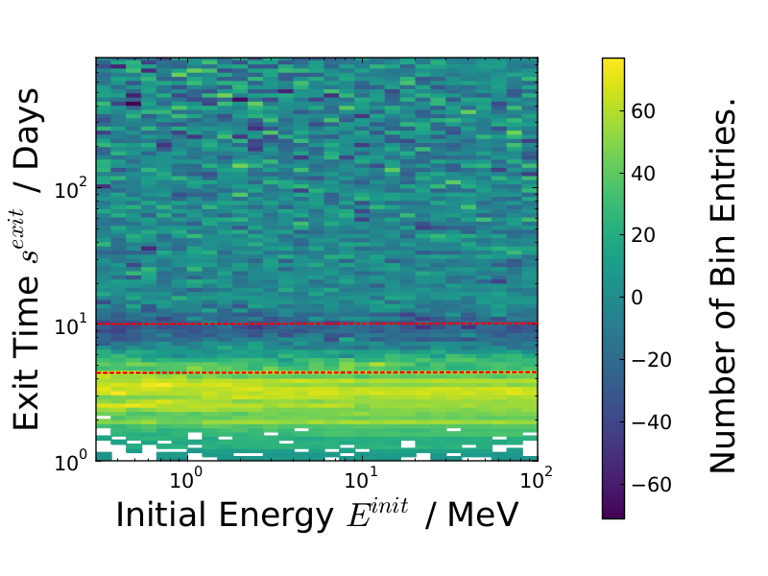}
            \caption[]%
            {
            }    
            \label{fig:influence_spectrum_vs_energy_meantime_compare}
        \end{subfigure} 
        \hfill        
        \begin{subfigure}[b]{0.49\textwidth}
            \centering
            \includegraphics[width=\textwidth]{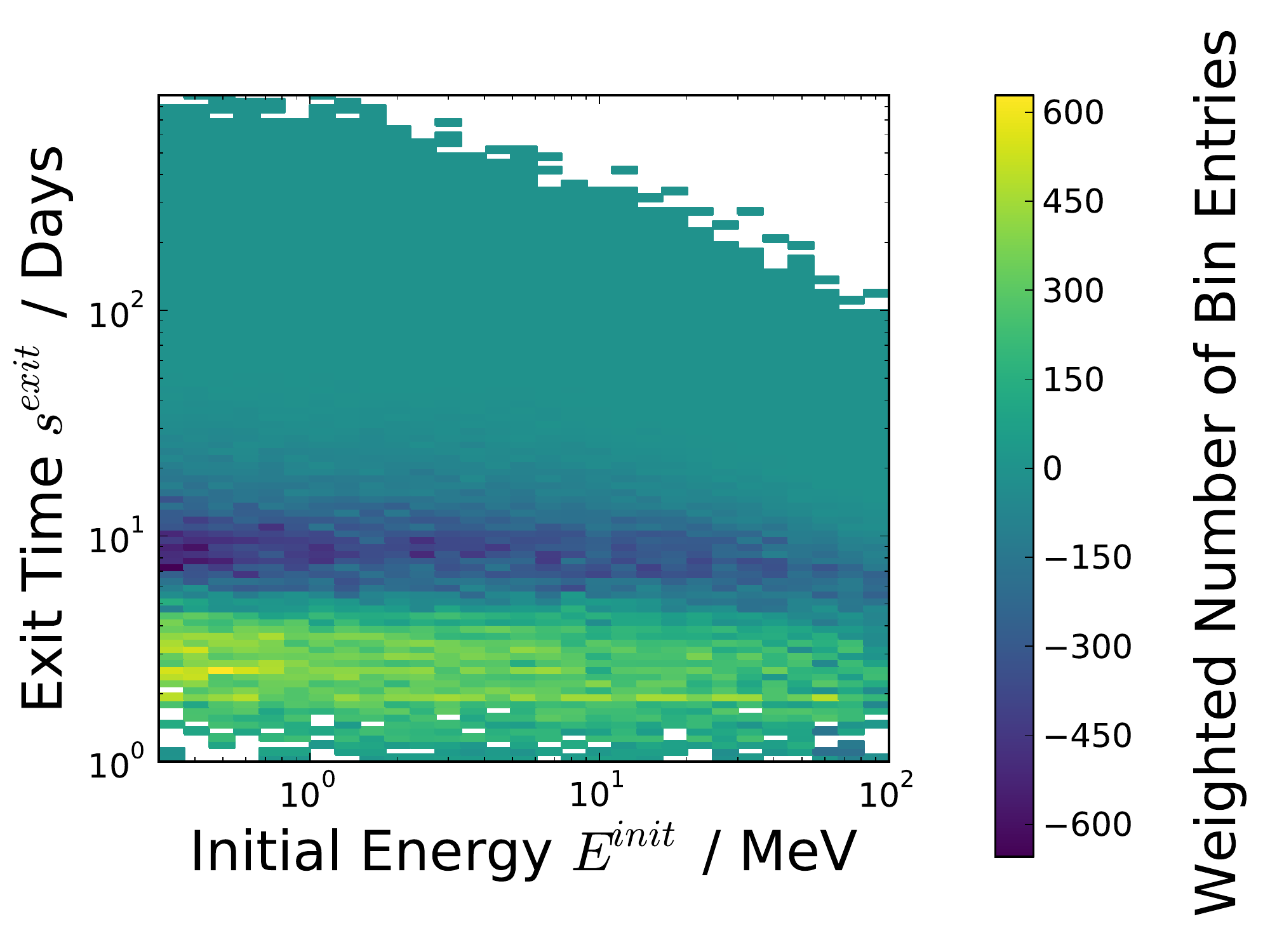}
            \caption[Network2]%
            {
            }    
            \label{fig:influence_spectrum_vs_energy_mpt_compare}
        \end{subfigure}        
    \caption{Binned distributions of exit times over the whole energy range dominated by Jovian electrons, 
    displayed similarly to Fig.~\ref{fig:influence_spectrum_vs_energy_flux}
    . Whereas the right panels show the exit time distributions weighted by their contribution according to Eq.~(\ref{eqn:mean_prop_time}), the left side displays the unweighted results of the simulation. The first row (Figures~\ref{fig:influence_spectrum_vs_energy_meantime_best} and \ref{fig:influence_spectrum_vs_energy_mpt_best}) shows the distributions for the case of good magnetic connection, and the second for the case of poor connection. Figures~\ref{fig:influence_spectrum_vs_energy_meantime_compare} and \ref{fig:influence_spectrum_vs_energy_mpt_compare} show the differences between the distributions for good and poor magnetic connection, highlighting the influence of magnetic connectivity on the simulation results}
    \label{fig:influence_spectrum_vs_energy_mpt}
\end{figure*}

Taking also Fig.~\ref{fig:influence_spectrum_vs_energy_mpt} into account in order to include the 
dependence of $\tau$ 
on the adiabatic energy changes
, the intensity sink 
(as indicated in red)
is confirmed 
over the whole energy range of interest.
The right panels show the histograms for the exit times at the Jovian boundary considering each to be of equal probability, while the left panels show their contribution to the residence time according to Eqn~\ref{eqn:mean_prop_time}. From top to bottom, first the case of good magnetic connectivity between observer and the source is shown, followed by the case of poor magnetic connectivity and the difference of the two sets of histograms. 
Similar to Fig.~\ref{fig:influence_spectrum_vs_energy_traj_compare}, which shows the difference of the phase-space trajectory distributions with respect to their corresponding exit energies, the difference of the phase-space trajectory distributions with respect to their corresponding exit times (Fig.~\ref{fig:influence_spectrum_vs_energy_meantime_compare}) also displays a significant maximum below the intensity sink (indicated in red). 
The comparison with Fig.~\ref{fig:influence_spectrum_vs_energy_mpt_compare} confirms that these trajectories indeed determine the residence time almost entirely. 
As discussed in detail above
, this 
further supports
that the population of pseudo-particles with trajectories dominated by the perpendicular diffusion process exit almost independently of the degree of magnetic connection between the source and the observer. 
However, the population corresponding to small exit times and low adiabatic energy gains seems to determine the resulting residence times, as it determines the resulting differential intensities. Since both small exit times and low adiabatic energy gains indicate very effective particle transport, it is reasonable to identify this population as being dominated by particles whose transport is governed by the parallel diffusion process as suggested by \cite{vogt2019}.

In the case of poor magnetic connection however, Fig.\ref{fig:influence_spectrum_vs_energy_meantime_worst} lacks the parallel diffusion-dominated population, but otherwise shows a similar shape. 
Again, the comparison with the corresponding contribution to the total differential intensity, as shown by Fig.~\ref{fig:influence_spectrum_vs_energy_mpt_worst}, suggests that the trajectories dominated by the perpendicular diffusion processes could be considered as being a 'diffusive background', 
since they neither change according to the degree of magnetic connection nor do they influence the physical results significantly.
This is due to the fact that the pseudo-particles' trajectories are the results of a Wiener Process as the mathematical representation of Brownian motion. Although the concept of parallel diffusion mimics the actual physical particle motion in the case of good connectivity and therefore shows the significance illustrated by Figs.~\ref{fig:influence_spectrum_vs_energy_meantime_best} and \ref{fig:influence_spectrum_vs_energy_mpt_best}, nevertheless the results as a whole have to be transformed according to Eq.~(\ref{eqn:diff_intensities}) in order to assign a physically 
significant
number of real particles to the pseudo-particles trajectories. This, as discussed in detail by \cite{vogt2019}, is of course equally applicable to the relation between the phase space trajectory exit times and the residence time as a result of applying Eq.~(\ref{eqn:mean_prop_time}).

The bottom panels of Fig.~\ref{fig:influence_spectrum_vs_energy_mpt} confirm these assumptions: As demonstrated by Fig.~\ref{fig:influence_spectrum_vs_energy_meantime_compare} and \ref{fig:influence_spectrum_vs_energy_mpt_compare} the 'diffusive background' disappears almost entirely if only the differences between the histograms for good and poor connectivity are shown. Therefore the suggested interpretation would be that the possibility to reach the Jovian boundary via a perpendicular diffusion-dominated trajectory does not depend on the magnetic connectivity between the boundary and the start or observational point. 
The fact that the diffusive background appears to be invariant with respect to the magnetic connection between the observer and the source, however, shows that this population of phase-space trajectories is dominated by perpendicular diffusion since otherwise it would not have reached the Jovian source as a boundary. Only 
with increasingly good connectivity will the probability that trajectories dominated by parallel diffusion reach the boundary increase. 

This effect is almost independent of the particle's energy. Although Fig.~\ref{fig:influence_spectrum_vs_energy_meantime_best} shows slightly shorter maxima of $s^{exit}$ for higher initial energies $E^{init}$, this effect is almost entirely limited to the population of trajectories which do not significantly contribute. The most prominent energy-dependent effect appears to be the increasing density of the distributions in Fig.~\ref{fig:influence_spectrum_vs_energy_mpt_best}, leading to more distinct maxima as also slightly visible in Fig.~\ref{fig:influence_spectrum_vs_energy_meantime_best} for the same exit times and initial energies, respectively. The fact that the intensity sink is present over the whole energy range further supports the interpretation of a 'diffusive background' and trajectories reflecting the physical behaviour of charged particles within the heliospheric magnetic field -- as indicated by the fact that, in case of good magnetic connection, only the latter population seems to contribute to the resulting differential intensity and residence times.

\section{Comparison of numerical and analytical approaches}
\label{sec:compare_analytical_numerical}

\begin{figure*}
        \centering
        \begin{subfigure}[b]{0.49\textwidth}  
            \centering 
            \includegraphics[width=\textwidth]{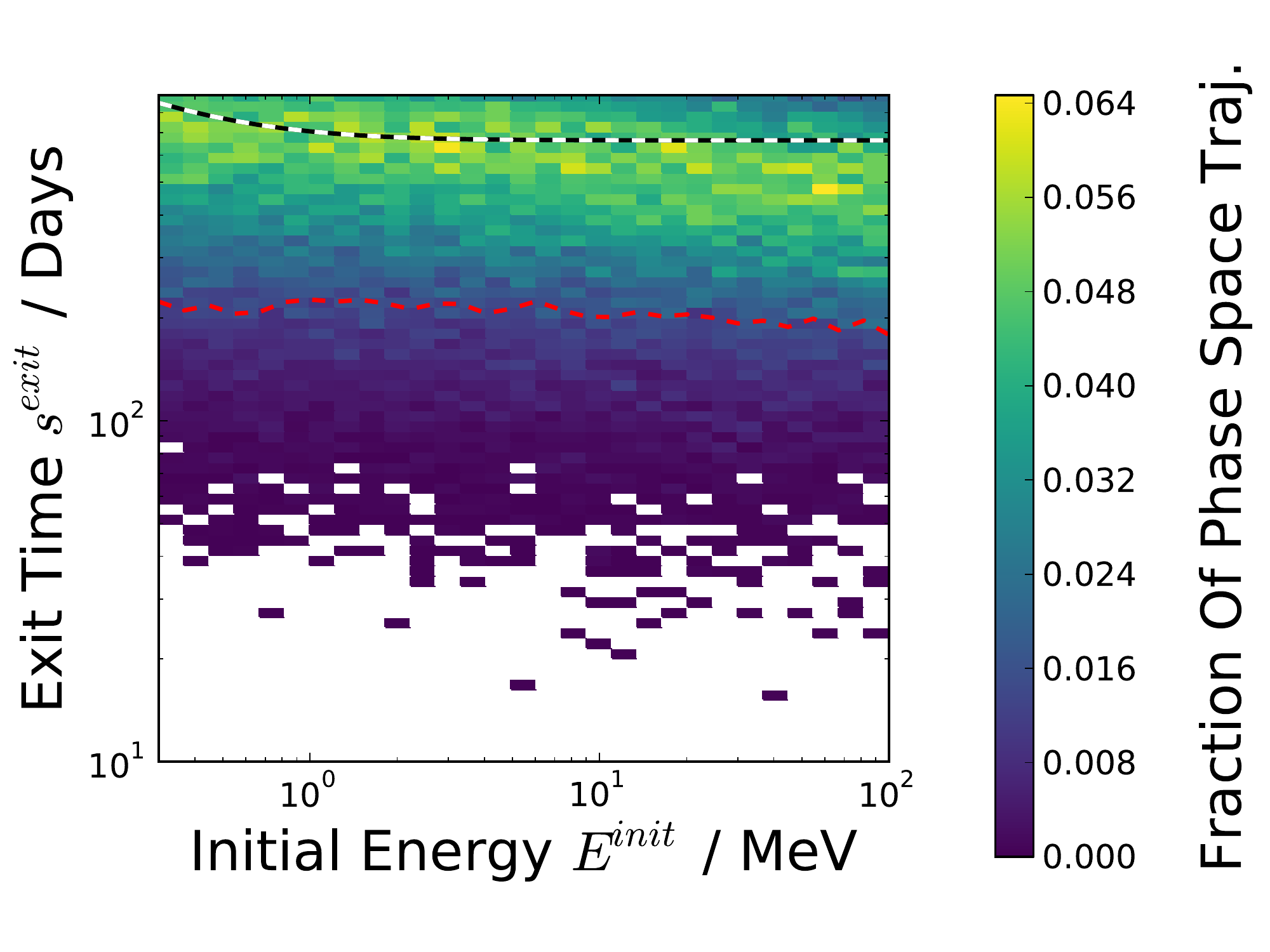}
            \caption{
            }   
            \label{fig:gal_mpt_meantime_simulation}
        \end{subfigure} 
        \hfill        
        \begin{subfigure}[b]{0.49\textwidth}
            \centering
            \includegraphics[width=\textwidth]{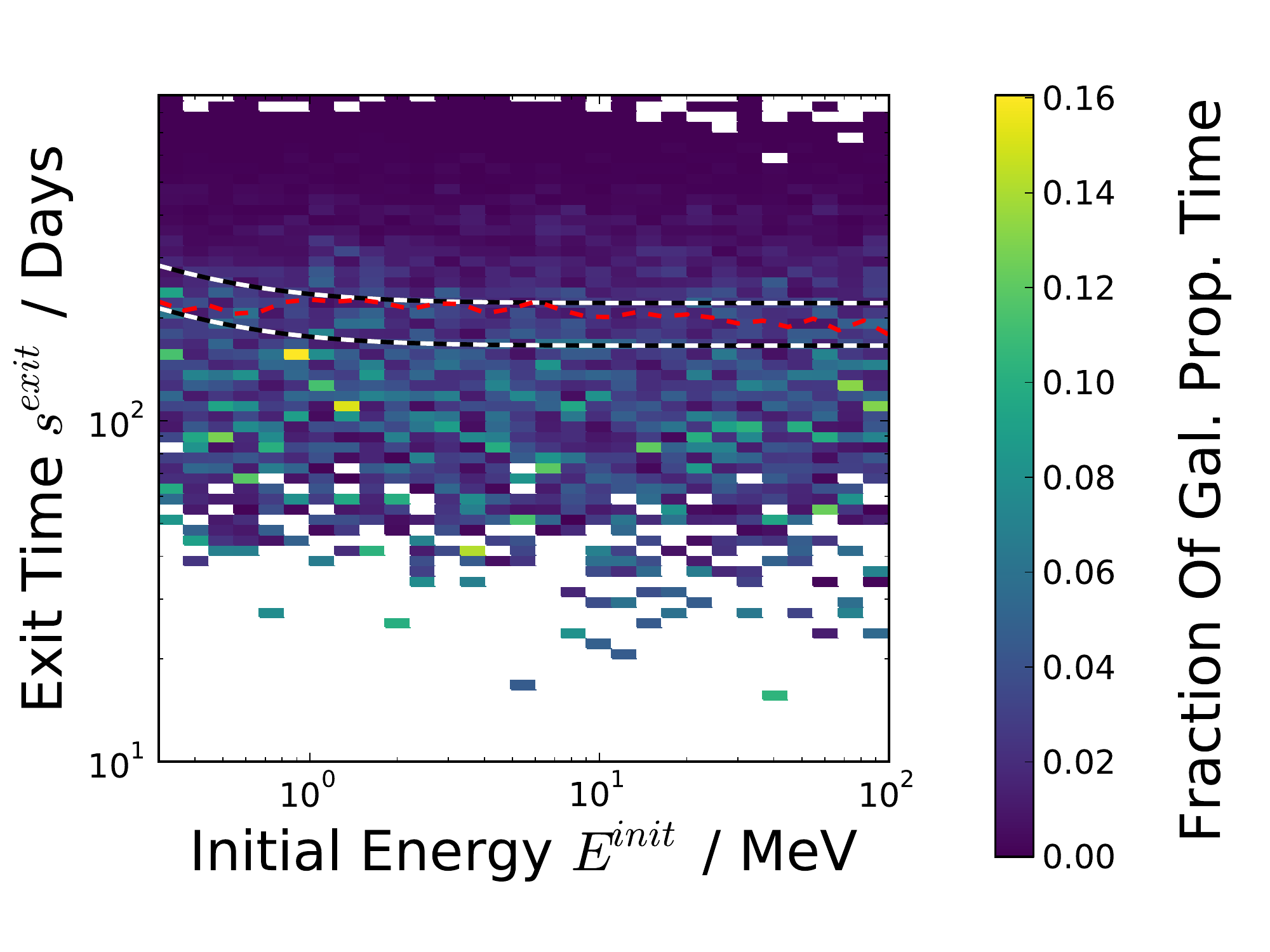}
            \caption{
            }
            \label{fig:gal_mpt_meantime_analytical}
        \end{subfigure}
        \caption{Similar to Fig.~\ref{fig:influence_spectrum_vs_energy_mpt} but for trajectories exiting at the outer boundary are shown, instead of the Jovian magnetosphere. The left panel shows the distribution of unweighted exit times $s^{exit}_i$ of trajectories exiting at the outer boundary. Note that the maxima are in loose agreement with the analytical estimation as suggested by \cite{parker1965} 
        given by the dashed white and black line}. The right panel shows the influence of weighting with Eq.~(\ref{eqn:mean_prop_time}) together with numerical solution indicated in red and two estimations of the revised analytical solution given by Eq.~(\ref{eqn:tau_analytical}). 
        \label{fig:gal_mpt_meantime}
\end{figure*}


\begin{figure}
            \centering
            \includegraphics[width=\columnwidth]{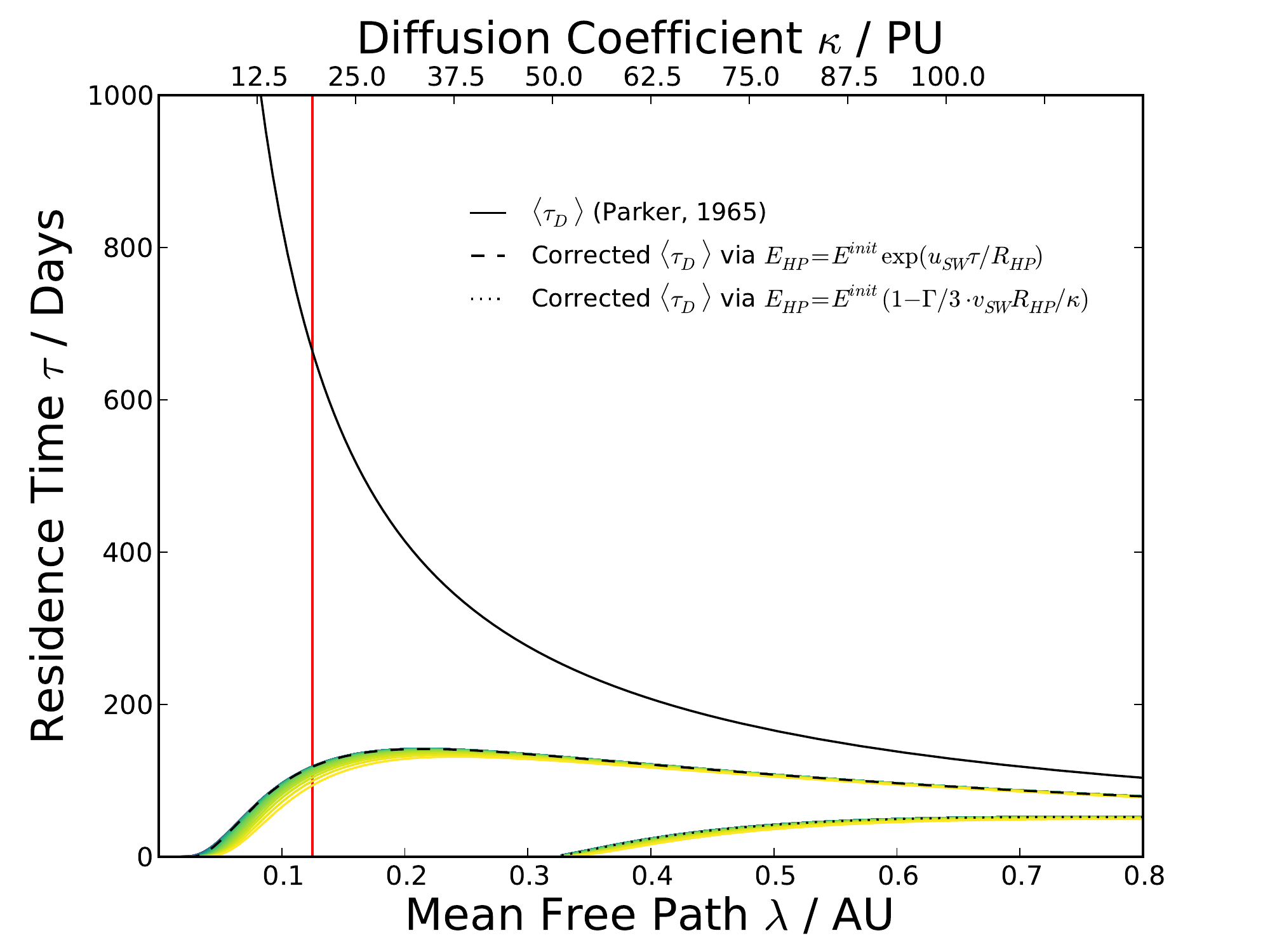}
   \caption{
   The analytical estimate of the \ac{GCR} residence time according to \cite{parker1965} with respect to the assumed mean free path $\lambda$ (bottom x-axis) and the resulting diffusion coefficient $\kappa$ (top x-axis). The two dashed lines show the corrected analytical estimates according to the estimations for the adiabatic energy loss by Eq.~(\ref{eqn:fract_energy_loss1}) and \ref{eqn:fract_energy_loss2}, respectively. The resulting residence times are shown, colour coded from blue to yellow, for the whole energy range of the Jovian spectrum (by 30 logarithmically-spaced representive energies) as a measure of the uncertainty.} 
   \label{fig:analytical_estimates}
\end{figure}

The large discrepancy between between the maxima of the distribution of the exit times $s^{exit}$ and the calculated residence times as displayed by the comparison of the left and right panels of Fig.~\ref{fig:influence_spectrum_vs_energy_mpt} seems to be caused by the mathematical nature of the diffusion approach. As already outlined by \cite{vogt2019} 
and discussed in Sec.~\ref{ssec:SDE_modelling} and especially Sec.~\ref{ssec:energy_dep} focusing on the role of adiabatic energy changes
, the solutions of the \ac{SDE} modelling 
approach (i.e. pseudoparticle trajectories)
have to be transformed into physically meaningful quantities 
(i.e. differential intensities)
by the means of the source spectrum 
as a boundary condition
. A probable consequence becomes apparent in 
comparison
to the analytical approximations of \cite{parker1965} and \cite{OGallagher1975}. 
Comparing these analytical estimations to the results of our approach to simulate the residence times shows a pattern of deviations similar to the deviations of the exit time distributions.

As the solution of the \ac{TPE} is a four-dimensional probability density function of time, space and rigidity (or energy, respectively), the residence time can be derived from it as the expectation value of the time as discussed at the beginning of Sec.~\ref{sec:residence_times} by means of Eq.~(\ref{eqn:expectation_value_time}). 
In order to solve the \ac{TPE} analytically, however, it has to be simplified. Therefore the assumptions are made 
by \cite{parker1965} (and subsequently for all other analytical estimations so far as well)
that  $\kappa$ and $u_{SW}$ are constant 
scalar values in time and space.
Furthermore it is assumed that the particles diffusion inwards are isotropic and that the particles propagate from the outer border of the heliosphere inward toward its center. 
Solving Eq.~(\ref{eqn:expectation_value_time}) under these assumptions yields a diffusion and a convection limit $\tau_D = R^2/(6\kappa)$ 
\citep[see][]{parker1965}'
and $\tau_C=R/u_{SW}$ 
\citep[see][]{OGallagher1975}
, respectively. For a more detailed description of the derivation see 
also
\cite{strauss2011}, and references therein. Evidently, both estimations are independent of the particle energy. The diffusion limit $\tau_D$, however, incorporates an indirect dependence via the overall average value of the diffusion coefficient $\kappa$. The assumption of a globally constant value of $\kappa$ 
(which does not depend on the radial distance nor differentiates between parallel and perpendicular diffusion processes)
of course is not realistic for the case of Jovian electrons, which have transport properties that depend strongly on whether parallel or perpendicular diffusion dominates. Therefore Galactic electrons have to be taken into account in order to compare these estimations with simulation results. 

For the purpose of these qualitative considerations, we used the galactic electron \ac{LIS} proposed by \cite{Potgieter2015} in order to apply Eq.~(\ref{eqn:mean_prop_time}) to the phase space trajectories exiting at the outer boundary equivalent to the way the Jovian residence times are calculated.
Although the same transport processes also govern the propagation of \ac{GCR} electrons,  their isotropic influx allows one to roughly combine parallel and perpendicular diffusion into one effective value of $\kappa$ according to e.g.\cite{parker1965}. Nevertheless the results of \cite{strauss2011b,Strauss2013}, who benchmarked their numerical estimation in case of \ac{GCR} electrons against the analytical diffusion limit $\tau_D$ by \cite{parker1965,Parker1966} and a combined estimation of $\tau=(1/\tau_D-1/\tau_C)^{-1}$, show in the Jovian electron case that values of $\lambda_{\parallel}=\lambda_{\perp}\approx1~$AU are required in order to obtain realistic values for the Jovian residence times, a result which is contrast to values successfully employed in modulation studies by e.g. \cite{Ferreira2001,Ferreira2001b, Kissmann2004,Sternal2011, strauss2011,Nndanganeni2016}, amongst others, as well as in contrast with what is expected from theory \cite[e.g.][]{EB2013,DempersEngelbrecht20,Shalchi20} and numerical test-particle simulations of diffusion coefficients \cite[e.g.][]{Minnie2007diff,hs16,Dundovic20}.
Please note that the increase of the mean free paths with radial distance has to be balanced with the fact that also the much less effective perpendicular diffusion has to be considered. Therefore, the assumption that $\lambda_{\parallel}=\lambda_{\perp}\approx1~$AU, which is almost an order of magnitude larger than the parallel mean free path at Earth orbit \cite[see, e.g.,][]{Bieber1994} and almost three orders of magnitude larger than the perpendicular mean free path \cite[see, e.g.,][]{Palmer1982,Giacalone98}, appears to be unrealistic.

This discrepancy is most likely caused by the fact that neither of the analytical estimates incorporate the adiabatic energy changes, and therefore 
to a degree
neglect the boundary conditions. As argued above in Sec.~\ref{ssec:adiabatic_cooling}, the \ac{TPE} is 
solved for an isotropic diffusion approach 
and a flat energy spectrum as input. This simplification appears to be problematic because it neglects the physical constraints of charged particle transport theory. 
Most significantly, it should be noted that
the diffusive processes incorporated into the \ac{TPE} describe stochastic processes associated with (pitch-angle) scattering at irregularities within the \ac{HMF}, as acknowledged by \cite{parker1965}. 
The effect of the source spectrum as the boundary condition included via the adiabatic energy changes is discussed in Sec.~\ref{ssec:SDE_modelling} and in great detail by \cite{vogt2019} and illustrated by Fig.~\ref{fig:simulation}. Summarised, and put in mathematical terms, the analytical estimates by \cite{parker1965}, \cite{OGallagher1975} and \cite{strauss2011b} are the temporal expectation value of the probability density function and not (as the boundary conditions are neglected) of the corresponding differential intensity.
In order to avoid the consequences of over-simplifying within the analytical estimates 
by neglecting the boundary conditions and neglecting realistic diffusion conditions
, it therefore seems to be necessary to incorporate an analytical estimate of the 
of the effect of
boundary conditions 
. 


Indeed,
Fig. \ref{fig:gal_mpt_meantime}
shows that the main effect of weighting the trajectories according to their contribution to the differential intensity determines the results for the Galactic population too. As one would expect, the analytical estimation 
of $\tau_d$ according to \cite{parker1965}
(dashed white and black line) in Fig.~\ref{fig:gal_mpt_meantime_simulation} 
is
in agreement with the unweighted distribution of exit times. The numerical estimation 
according to \cite{vogt2019} 
, however, as shown by the red dashed line, strongly deviates in this case, 
similar to the case of the Jovian electron population
. These findings 
confirm
our interpretation 
of the significance of including the boundary conditions via the 
adiabatic energy changes in order to transform the mathematical solution of the \ac{TPE} into a physical and observationaly meaningful one. 
Due to the adiabatic energy changes of particles, they are not detected with the energy they had at their source. In order to calculate the residence time of a particle detected with $E^{init}=6~$MeV at the observational point one therefore would have to assume an energy of $E^{exit}=E^{init}+\Delta E$ at the source with $\Delta E$ being the average adiabatic energy change of the particle population. Since these two energies correspond to different intensities if the spectrum is not assumed to be flat (
as is assumed here on the basis of observations
, see the discussion in Sec.~\ref{ssec:adiabatic_cooling}), the adiabatic energy changes can be transformed into intensity changes within this mathematical framework. 
Neglecting the influence of adiabatic energy changes would implicitly assume an energy-independent source spectrum.

The approach to 
estimate
the characteristic energy change as being proportional to the initial energy is supported by \cite{parker1965}, who assumed the energy loss after a time $t$ as being given by 
\begin{equation}
    \label{eqn:fract_energy_loss1}
    E(t)\approx E_0\exp(-tu_{SW}/R_{HP})
\end{equation} and that $t$ is not energy-dependent 
Solving a simplified \ac{TPE} accordingly,
this approach leads to an average fractional energy loss of
\begin{equation}
    \label{eqn:fract_energy_loss2}
    \left\langle\frac{E^{exit}}{E^{init}}\right\rangle= \left(1-\frac{\Gamma}{3}\frac{Ru_{SW}}{\kappa}\right)^{-1} \hspace{5mm}\mbox{with}\hspace{5mm} \Gamma=\frac{E^{exit}+2E^{rest}}{E^{exit}+E^{rest}}
\end{equation}
as derived by \cite{Parker1966}. As pointed out by e.g. \cite{strauss2011b}, this expression is only valid for a constant stream of in-flowing particles with a globally constant mean free path and diffusion tensor, respectively. 
Fig.~\ref{fig:analytical_estimates} shows the analytical estimate of the diffusion limit $\tau$ alongside its corrected value according to both estimates for the adiabatic energy changes, Eqns.~\ref{eqn:fract_energy_loss1} and \ref{eqn:fract_energy_loss2}. As outlined above 
in the second paragraph of this section
we therefore assumed the analytical estimate of the effect of the boundary conditions as being a function of the ratio of the source spectrum intensity $j_{LIS}(E)$ at the initial energy $E^{init}$ and at the exit energy $E^{exit}$, given by:
\begin{equation}
    \label{eqn:corrected_analytical_estimate}
        \langle\tau_D\rangle=\frac{R^2}{6\kappa}\cdot \left\langle\frac{j_{LIS}(E^{init})}{j_{LIS}(E^{exit})}\right\rangle
\end{equation}
A limitation of this approach is that only the diffusive limit $\tau_D$ is considered. The approach by \cite{strauss2011b}, which also considers the outward convection of the particles, is not defined for the small values of $\lambda$, and therefore of $\kappa$, that we need to apply for Jovian electrons as shown and discussed in Sec.~\ref{ssec:setup} and \cite{vogt2019}. Nevertheless the slightly lower values of $\tau_D$ (see \cite{strauss2011b} for detailed discussion) appears to be in agreement with the exit times $s^{exit}$ as displayed in Fig.~\ref{fig:gal_mpt_meantime_simulation}. 

As illustrated by Fig.~\ref{fig:analytical_estimates}, the explicit analytical solution for the fractional adiabatic energy loss given by Eq.~(\ref{eqn:fract_energy_loss2}) is also not physically usefully defined due to the unrealistic assumption made in its derivation that $R_{HP}/\kappa \ll 1$
, given that the Voyager mission confirmed that $R_{HP}\approx 120~$AU \citep[see e.g.][]{Gurnett2013,Krimiges2013, Stone2019} and that our results for $\lambda_{\parallel/\perp}$ suggest significantly smaller values for $\kappa$
. 
Therefore, as indicated by the wider range of definition in Fig.~\ref{fig:analytical_estimates}, we 
have to%
chose to apply Eq.~(\ref{eqn:fract_energy_loss1}) in order to estimate the effect of the boundary conditions on the analytical estimate for the residence time
, otherwise our value for $\lambda$ would be outside the range covered by the assumptions.
As the energy $E(t)$ in Eq.~(\ref{eqn:fract_energy_loss1}) is given as depending on the time $t$ and the energy $E_0$ with which the particle enters the heliosphere, Eq.~(\ref{eqn:fract_energy_loss1}) has to be rewritten according to the time-backward approach, and we therefore have to compare our analytical estimate with:
\begin{equation}
\label{eqn:estimation_delta_E}
E(t)\approx E^{init}\exp(tu_{SW}/R_{HP})\mbox{.}
\end{equation}
Since we are interested in the total average energy gain for \ac{GCR} electrons we substitute $t$ with the diffusion limit $\tau=R_{HP}^2/(6\kappa)$ and obtain
\begin{equation}
\label{eqn:estimation_delta_E2}
E^{exit}(\tau_D)\approx E^{init}\exp\left(\frac{R_{HP}u_{SW}}{6\kappa}\right)\mbox{.}
\end{equation}
As indicated by Eq.~(\ref{eqn:adiabatic_cooling}) the adiabatic energy changes are more effective whenm the radial distance at chich the particle finds itself is lower. Therefor, 
we assume the value of $\kappa$ as corresponding to 
the value of the parallel mean free path
$\lambda_{\parallel}=0.125~$AU at Earth orbit (marked in red in Fig.~\ref{fig:analytical_estimates}). 
Furthermore, the analytical estimates cannot distinguish between perpendicular and parallel diffusion. Therefore $\kappa$ has to incorporate not only the effective diffusion parallel to the magnetic field but also the diffusion perpendicular to the field lines, which is less effective by almost two orders of magnitude as shown by the value of $\chi$ in Tab.~\ref{tab:simulation_parameters}. Due to the much longer path length of the Parker spiral in comparison with the radial distance, the more ineffective perpendicular transport cannot be ignored. Our assumptions are 
supported by Fig.~\ref{fig:gal_mpt_meantime_simulation} as it shows that the similarly calculated diffusion limit $\tau=R_{HP}^2/(6\kappa)$ is in agreement with the simulation results. Despite being located at the lower edge of the definition range and the uncertainties implicit to the need to assume an effective global value of $\kappa$, Fig.~\ref{fig:analytical_estimates} supports the 
qualitative
notion that the effect of the boundary conditions can be estimated as being a factor of $1/3$ to $1/4$ as 
indicated 
by the black and white dashed lines in Fig.~\ref{fig:gal_mpt_meantime_analytical}. 

We can demonstrate that this
analytical estimation shifts to the same range of values as suggested by our numerical estimation 
given by Eq.~(\ref{eqn:mean_prop_time}) and indicated by the dashed red line.
Whereas a more detailed treatment of how to define and derive a characteristic energy change is beyond the scope of this study, the approach sketched herein should be sufficient to qualitatively strengthen our point as to how diffusion and adiabatic energy changes are physically interlinked with each other within the theoretical treatment of charged particle transport. As described by \cite{vogt2019} the approach to model diffusion via a Wiener process does not track the particle motion itself, leading to mathematical remnants such as 
the
'diffusive background' as discussed in Secs.~\ref{ssec:SDE_modelling} and \ref{ssec:energy_dep}, which then have to be transformed into physically 
useful
solutions by convolution with the source spectrum. The comparison between analytical and numerical estimations, however, suggests that these considerations apply to diffusion as a mathematical model to treat particle motion in general, rather than just to its numerical implementation by means of \acp{SDE}. 

\section{Corotational effects as possible measure}
\label{sec:corotation}

\begin{figure*}
    \centering
    \includegraphics[width=\textwidth]{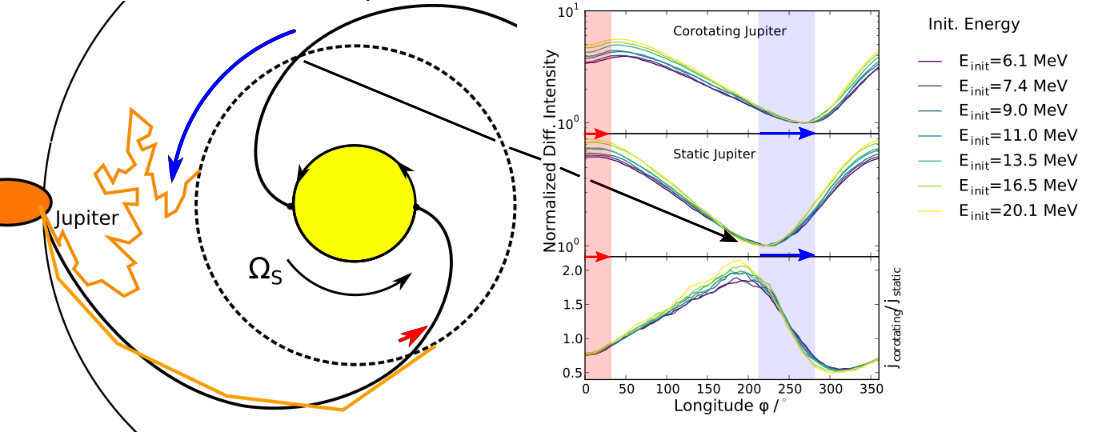}
    \caption{
    Differential intensities of Jovian electrons for different initial energies $E^{init}$.The top panel shows the longitudinal variation due to varying connection for a co-rotating Jovian source, while the middle panel shows the case of a static Jupiter. The bottom panel shows the ratio of the results of the two approaches. Right panel taken from \cite{vogt2019}}
    \label{fig:Jovian_corotation}
\end{figure*}

One way to examine these considerations is 
to utilize
the corotational effects as described by \cite{vogt2019}. 
As described in Sec.~\ref{ssec:setup}, the \ac{HMF} is considered only by its geometry and not directly implemented itself. Because also the Solar rotation is neglected by the simulation setup, the magnetic connection between the Jovian source and the observer does not change throughout the simulation - or put more explicitly, for the duration of each simulation run the observer is located on the same Parker magnetic field line. In order to investigate the effects caused by these simplifications \cite{vogt2019} implemented the effect of the Solar rotation (and therefor the co-rotation of the \ac{HMF}) by considering the longitudinal motion of the Jovian source relative to the static \ac{HMF} geometry.
Comparing simulations over the whole longitudinal range with both a static Jovian source and a co-rotating one reveals a significant deviation of about $50^{\circ}$ from the nominally best-connected longitude
, an effect even larger in case of poor magnetic connection between observer and the Jovian source. Utilizing simulation results from \cite{vogt2019},  Fig.~\ref{fig:Jovian_corotation} sketches the physical mechanism behind this.
Two different effects 
determine
the magnetic connection between Jupiter and a possible observer:
\begin{enumerate}
    \item The synodic period, in case of Earth the characteristic $\approx 13$ months first discussed by \cite{McDonald72}.
    \item The corotation of the \ac{HMF} due to the Sun's rotational period of $\Omega_S\approx 27$ days 
\end{enumerate}
The synodic corotation 
of $\approx 13$ months
bears more or less no detectable effect since the expected angular shift between Earth and Jupiter in case of good magnetic connectivity is about $\Delta \phi \lessapprox 4^{\circ}$. The angular speed of the \ac{HMF} 
caused by the Solar rotation with $\Omega_S\approx 27$ days,
however, suggests a much more significant angular deviation of $\Delta \phi \gtrapprox 50^{\circ}$ 
confirmed
via simulation 
by \cite{vogt2019}. 
The corresponding relative motion between the \ac{HMF} and the Jovian source thereby is implemented as a co-rotation of Jupiter by an additional term in the longitudinal \ac{SDE} as given by Eq.~(\ref{eqn:sde_derived}). A similar approach is taken by \cite{Ferreira2002Phd}.
The deviation of the maximum intensity from the nominally best-connected longitude caused by this adjustment of the simulation setup
is indicated by the red shaded shift in Fig.~\ref{fig:Jovian_corotation}. As illustrated in the left panel of Fig.~\ref{fig:Jovian_corotation}, the reason for this shift is the following: Due to the Sun's angular rotation, the nominal Parker-like \ac{HMF} frozen into the radially outward streaming Solar wind also rotates with the same angular velocity. During the propagation of Jovian electrons (indicated for good and poor connections by the orange trajectories in Fig.~\ref{fig:Jovian_corotation}), a Parker-spiral connecting the Jovian source with the Earth orbit (dashed black circle) would change its longitudinal position, as indicated by the red arrow. Since the Jovian electron could follow the path of the nominal Parker-spiral by the more effective processes categorized as parallel diffusion (indicated by the larger step sizes in Fig.~\ref{fig:Jovian_corotation}), the residence times in the case of good magnetic connection are small and only cause small corotation effects. For the case of poor magnetic connection, however, perpendicular diffusion dominates the propagation processes of Jovian electrons that happen to reach the observer at Earth orbit, and therefore leads to much longer residence times. As indicated by the blue arrow and the blue shaded area in Fig.~\ref{fig:Jovian_corotation}, respectively, the corresponding angular shift of the mininum of the 13 month periodicity of Jovian electron intensity is expected to be much larger. However, due to the longer residence times, the fluctuations in both Solar wind speed and the \ac{HMF} can be expected to significantly interfere with this effect. But even in case of good magnetic connection, the angular shift due to the corotation is expected to be
large enough to be possibly resolved by spacecraft data and therefore could serve as an (indirect) measure for the residence time via
\begin{equation}
\label{eqn:angular_sep_tau}
    \tau=\frac{\Delta \phi}{\Omega_{S}}
\end{equation}
with $\Omega_{S}$ being the Sun's angular speed. Although conceptually simple, this approach brings about several difficulties concerning the data analysis it requires. Despite being generally difficult to detect \citep[see e. g.][amongst others]{Heber2005}, electron counting rates are often highly influenced by \ac{SEP} events and \acp{CIR}. Furthermore the \ac{HMF} as defined by \cite{Parker1958} and used by the modelling setup \citep[see Sec.~\ref{ssec:setup} as well as][]{Dunzlaff2015,vogt2019} assumes a globally constant solar wind velocity, which is a strong simplification on the longer timescales involved, even during periods of low solar activity. 

\begin{figure}
            \centering 
            \includegraphics[width=\columnwidth]{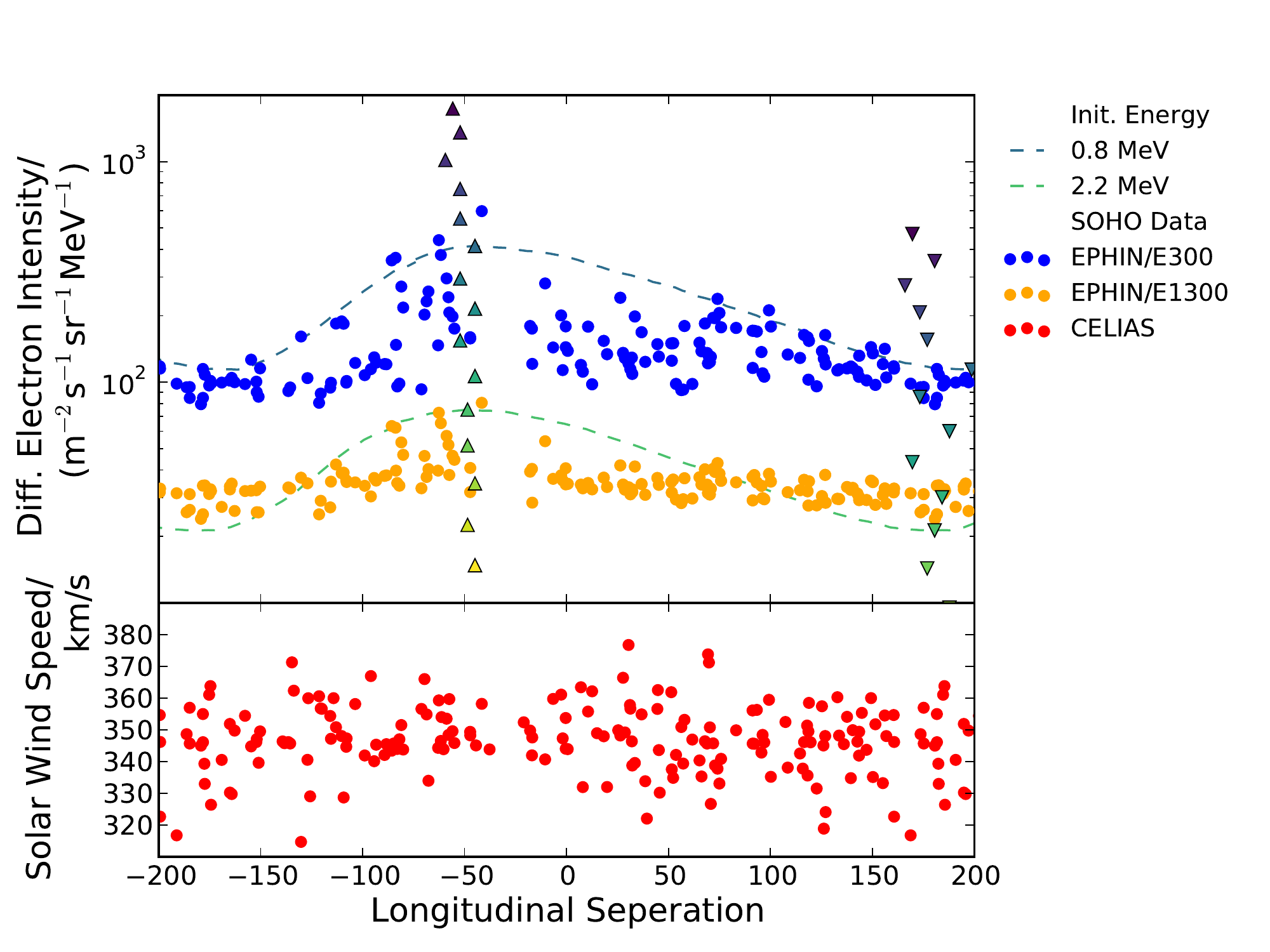}
   \caption{The effect of corotation as shown by Figure~\ref{fig:Jovian_corotation} as detected by SOHO-EPHIN. Whereas the upright and down right triangles mark the positions of the maxima and minima of the simulated Jovian fluxes as shown in the upper panel of Figure~\ref{fig:Jovian_corotation}, the two electron channels are shown in blue and orange as daily averages sampled over four synodic periods during the 2006-2011 Solar minimum. The simulation results corresponding to initial energies equivalent to those of the electron channels are shown 
   to match
   to the envelope of the data. In the lower panel, the corresponding SOHO-CELIAS data points are given for the daily averages of the Solar wind speed. Both panels show their data with respect to the longitudinal separation to the nominal point of best magnetic connection.} 
    \label{fig:corotation_ang_seperation}
\end{figure}

Fig.\ref{fig:corotation_ang_seperation} therefore shows SOHO-EPHIN electron data covering four synodic periods between 2006-2011, plotted with regard to their longitudinal separation from the nominal angle of best connectivity. The \ac{HMF} for this matter was assumed to be frozen into a solar wind with a radial speed of $u_{SW}=400~$km/s. In order to constrain the database to the boundaries given by means of the simulation setup, daily differential electron intensities were only included if all hourly averages of the solar speed by SOHO-CELIAS were within the range of $[300,400]~$km/s. The remaining daily averages of solar wind speeds are plotted in the lower panel of Fig.~\ref{fig:corotation_ang_seperation}. The corresponding electron intensities of the E300 (blue) and E1300 (orange) channels of the EPHIN instrument still show a large range of variation but an envelope in agreement with the simulation results. This can be understood by way of how solar wind speed variations influence the propagation between the Jovian source and the observer. By choosing only days with solar wind speeds with hourly averages 
not bigger than
$u_{SW}=400~$km/s, we assure that these variations change the longitude of best connectivity due to decreases of the solar wind speed, resulting in a shift to the right seen in Fig.~\ref{fig:corotation_ang_seperation}. The effect of these shifts of connectivity is a change of the corresponding electron intensity. 

If the observer were well connected to the Jovian source by a \ac{HMF} corresponding to $u_{SW}=400~$km/s, a decrease of the solar wind speed would lead to a loss of connectivity and the Jovian electron intensity would decrease likewise. Thus only the upper envelope of the data is comparable to the simulations with an undisturbed \ac{HMF}. 
Another reason to limit the focus on the envelope as a test of principle is the fact that the Solar wind takes several days to propagate across the $4~$AU between Earth and Jupiter. Thus to be in complete agreement with the simulation we would have needed data corresponding to days of stable Solar wind conditions. Since Fig.~\ref{fig:corotation_ang_seperation} already displays five years of data during the most stable conditions measured by SOHO-EPHIN, this simply is a problem of statistics. But
as shown in the upper panel of Fig.~\ref{fig:corotation_ang_seperation}, both of the electron channels of EPHIN meet the theoretical and computational expectations 
within
the restrictions described. Whereas the simulation results are in agreement, both qualitatively and quantitatively, with observations from the E300 channel, the intensities provided by E1300 are higher than expected. There are two possible reasons for this divergence. In order to calculate a representative energy for each channel, the detector responses \citep[see][]{kuehl2013} have to be convoluted with the electron spectrum. Apart from these technical difficulties, the Galactic component could also contribute more significantly at the measured energies than expected, despite the results of \cite{Nndanganeni2018} who suggest that there is no significant Galactic electron contribution up to measured energies of $\approx 25~$MeV. Additional to these possible statistical causes, another reason for the deviation could be the fact that, due to smaller fluxes, the unstable \ac{HMF} conditions as described above could in general have a larger impact on the data. 

Nevertheless the fact that the envelopes of both electron channels plotted in Fig.~\ref{fig:corotation_ang_seperation} are in agreement with both the simulation results and the theoretical expectation can be considered as a 
confirmation
of concept. Despite the uncertainties tied to the fluctuation in the propagation conditions, this result serves as an encouragement to investigate this approach further when future spacecraft missions may provide data with improved statistics.

\section{Discussion and conclusions}
\label{sec:discussion}

This paper investigated the energy dependence of residence times, focusing on the effect that adiabatic energy changes have on the mathematical implementation of diffusion. We discussed how adiabatic energy changes and and simulation times are related within the framework of \ac{SDE} modelling. These considerations in Sec.~\ref{ssec:energy_dep} were taken into account in order to investigate how different initial energies $E^{init}$ influence the estimated residence times according to Figs.~\ref{fig:influence_spectrum_vs_energy_mpt}. The significant features of the results (both weighted and unweighted for good and poor magnetic connection) were used to further explore how the mathematical results are related to the physically observable quantities. 

This question initially lead to the suggestion to calculate residence times as proposed by \cite{vogt2019} and outlined in Sec.~\ref{sec:residence_times}, but appears to be of broader significance. As we showed by 
comparing
the numerical and analytical approaches concerning the residence times of \acp{GCR}, the prior analytical approaches to estimate residence times are equivalent to taking the simulation or exit times $s^{exit}$ as a measure for $\tau$. Therefore this study provides 
evidence
to include representative adiabatic energy changes into the analytical approach. As discussed in Secs.~\ref{ssec:energy_dep} and \ref{sec:compare_analytical_numerical}, it has long been established that the convolution with the source (or equivalent boundary conditions in a non time-backward \ac{SDE} approach) is necessary in order to transform the results of a diffusion equation 
describing
Brownian motion into a physically 
relevant
solution of a more complex stochastic process. By means of Eq.~(\ref{eqn:corrected_analytical_estimate}) we suggest a simple but sufficient approximation which appears to reproduce the simulation results and could further serve as a more reliable first estimate for residence times 
when 
simulation results are not available. 

Utilizing the fact that the residence times result in a longitudinal shift of the best magnetic connection to the Jovian source due to relative orbital motions, as discussed in Sec.~\ref{sec:corotation}, we are able to show via Fig.\ref{fig:corotation_ang_seperation} that our estimations are in a general agreement with the observable longitudinal dependence of electron intensities. Since the estimation of residence times is connected to adiabatic energy changes (and therefore spectral modulation) and as well as the transport parameters within the \ac{TPE}, this offers a further possibility to constrain charged particle transport simulations. An example of is is the investigation of \acp{CIR} and their effect on particle propagation. The questions as to whether and how the considerations and results of this study can be generalized and applied to focused transport equations describing \ac{SEP} events, amongst others, is beyond the scope of this work, but appears to be a promising task for subsequent research.

%
\begin{acknowledgements}
This work is based on the research supported in part by the National Research Foundation of South Africa (Grant Number 111731). Opinions expressed and conclusions arrived at are those of the author and are not necessarily to be attributed to the NRF. 
KH gratefully acknowledges the support from the German Research Foundation (Deutsche Forschungsgemeinschaft, DFG) priority program SPP 1992 “Exploring the Diversity of Extrasolar Planets" through the project HE 8392/1-1, and further likes to thank ISSI and the supported International Team 464 (\href{https://www.issibern.ch/teams/exoeternal/}{ETERNAL}).
\end{acknowledgements}
\bibliographystyle{aa}
\bibliography{refs,add}
\end{document}